\documentclass[letter,12pt]{article}
\pdfoutput=1 

\usepackage{jheppub} 

\usepackage[utf8]{inputenc}

\usepackage{epsfig}
\usepackage{amssymb}
\usepackage{amsfonts}
\usepackage{amsmath}
\usepackage{titlesec}
\usepackage{yfonts}
\usepackage{booktabs}
\usepackage{mathrsfs}
\usepackage{amssymb}
\usepackage{calligra}

\usepackage{graphicx}

\usepackage{mathtools}

\usepackage{adjustbox}
\usepackage{stmaryrd}

\usepackage{arydshln}

\usepackage{ dsfont }

\usepackage[russian,english]{babel}
\usepackage[T1,T2A]{fontenc} 

\usepackage[utf8]{inputenc}
\usepackage{amsmath}
\usepackage{calc}

\newcommand{\x}{\rm x}
\newcommand{\y}{\rm y}
\newcommand{\chiralalgebra}{{\vphantom{\overline{\rm Vir}}{\rm Vir}}_{1/2}^{\otimes n} \times \overline{\rm Vir}_{1/2}^{\otimes n}}

\usepackage{accents}

\newcommand{\kket}[1]{\left|\!\left| #1 \right\rangle\!\right\rangle}

\def\ps{\mathfrak{\varphi}}

\newcommand{\avg}[1]{\langle\!\langle #1 \rangle\!\rangle}

\def\DD{{\sf D}}

\def\VTQFT{${\mathbb V}_{1/2}^n$TQFT}

\def\H{{\mathcal H}}

\def\C{{\mathcal C}}
\def\D{{\mathcal D}}
\def\I{{\mathcal I}}
\def\T{{\mathcal T}}

\def\Im{ \hbox{\rm Im}}

\newcommand{\Z}{\ensuremath{\mathbb Z}}
\newcommand{\R}{\ensuremath{\mathbb R}}

\newcommand{\Wh}{\mathcal{W} {\calligra h}}

\usepackage{amsmath}

\usepackage{ upgreek }
\usepackage{physics}

\def\bea{\begin{eqnarray}}
\def\eea{\end{eqnarray}}

\mathchardef\mhyphen="2D

\title{A solvable model of 3d quantum gravity}

\author{Anatoly Dymarsky${}^{a,b}$} 
\affiliation{${}^a$ School of Natural Sciences, \\  Institute for Advanced Study, \\ 1 Einstein drive, Princeton, NJ,  08540, USA\\}

\affiliation{${}^b$ Department of Physics and Astronomy, \\ University of Kentucky,\\ 505 Rose Street, Lexington, KY,  40506, USA\\}
\emailAdd{a.dymarsky@uky.edu}

\abstract{
We consider a model of 3d quantum gravity defined by $n$ copies of a rational Virasoro TQFT with central charge $1/2$, summed over all 3d topologies.  This theory is holographically dual to an ensemble of all 2d CFTs with central charge $c=n/2$ and chiral algebra that includes ${\rm Vir}_{1/2}^{\otimes n}$. We perform the sum over topologies and evaluate the 
partition function of the  bulk theory. We then confirm the holographic duality by matching it to the boundary ensemble for small $n$. We proceed to consider the limit of a large central charge, in which the bulk theory simplifies and condenses to an Abelian phase. In this regime, the model  manifests many features expected in semiclassical 3d quantum gravity. In particular, inclusion of all 3d topologies in the bulk sum cures the negativity of the density of states evaluated by the torus partition function. The model also exhibits a Hawking-Page transition, an exponentially suppressed wormhole amplitude, and provides a toy example of the holographic code. We discuss these aspects in detail and conclude with lessons for semiclassical quantum gravity. 

}

\begin{document} 
\maketitle
\flushbottom

\section{Introduction}
\label{sec:intro}
Three-dimensional quantum gravity with negative cosmological constant is a much-studied theory with many important developments to date \cite{Brown:1986nw,ACHUCARRO198689,Witten:1988hc,Banados:1992wn,Witten:2007kt,Maloney:2007ud,Hartman:2014oaa,Cotler:2020ugk,Chandra:2022bqq}. At the same time it is still not clear how to define it beyond the semiclassical limit. One way to define 3d quantum gravity non-perturbatively would be via holographic correspondence, by starting from a suitable two-dimensional CFT. This  approach encounters an immediate obstacle: there are no known large central charge, large spectral gap   CFTs that would be dual to semiclassical gravity. Known theories are dual to various strongly coupled models of gravity which do not teach us much about the semiclassical regime. 
A related effort to define 3d gravity as dual to an ensemble of all large-$c$ CFTs runs into its own problems, with the corresponding ensemble being notoriously difficult to define \cite{Belin:2025qjm}.

Given the topological nature of 3d gravity, a computationally powerful approach is offered by a reformulation of gravity in terms of a topological quantum field theory. We collectively refer to such theories as Virasoro TQFT 
\cite{Verlinde:1989ua,Kashaev:1998fc,Chekhov:1999tn,Teschner:2003at,Teschner:2010je,Mikhaylov:2017ngi, Collier:2023fwi,Collier:2024mgv,Hartman:2025cyj,Hartman:2025ula}.
This approach, too, is not without difficulty. While Virasoro TQFT is well-defined and matches semiclassical gravity for the ``on-shell''  topologies (hyperbolic manifolds), for ``off-shell'' topologies  (non-hyperbolic manifolds) it yields ill-defined results. This is a major obstacle, since the contribution of the off-shell topologies is necessary to make the 3d gravity well-defined \cite{Benjamin:2020mfz,Maxfield:2020ale}. 

In this paper we propose and solve a model that lies between the holographic and Virasoro TQFT approaches and hence has the potential to address some of the challenges outlined above. 
The model is well-defined both as a 3d bulk theory and in terms of the holographically dual boundary ensemble and exhibits qualitative features characteristic of semiclassical quantum gravity. The model is defined within the framework of TQFT gravity, by summing a topological quantum field theory  over all 3d manifolds with a set boundary \cite{Dymarsky:2024frx}. 
The TQFT in question is taken to be $n$ copies of the ``rational'' Virasoro TQFT with the value of central charge $c=1/2$ (combined with another $n$ anti-chiral copies).
For brevity we will refer to resulting model as \VTQFT\, gravity. 
The dual ensemble  includes all rational CFTs  with the central charge $c=n/2$ and the chiral algebra $\chiralalgebra$ or an extension thereof. The   sum over topologies that defines the bulk theory, as well as the details of the boundary ensemble are elucidated in section \ref{sec:sumovert}. 

\VTQFT\, gravity is a rational cousin of the Virasoro TQFT, completed with the prescription to sum over topologies.  It  is closely related to an earlier attempt to formulate  the gravity dual of the Ising CFT \cite{Castro:2011zq}, as well as the ``$U(1)$-gravity'' dual to the ensemble of all Narain theories \cite{Maloney:2020nni,Afkhami-Jeddi:2020ezh}, and followup works \cite{Ashwinkumar:2021kav,Meruliya:2021utr,Meruliya:2021lul,Raeymaekers:2021ypf,Aharony:2023zit,Angelinos:2025mjj}. The conceptual difference between the present model and these earlier examples is that the bulk sum is not limited to handlebodies but includes all possible 3d manifolds. Evaluating the bulk sum explicitly is the main technical result of this work.

The result of this calculation is as follows. 
The partition function of \VTQFT\, gravity -- averaged partition function of the boundary ensemble -- can be expressed as a sum over some auxiliary binary codes of a particular type, see \eqref{TPF} when  the boundary is a torus. There is no closed-form expression for finite $n$. The answer further simplifies in the large central charge limit $n\gg 1$, when the bulk theory ``condenses'' to an Abelian phase. In this phase it is described by the TQFT gravity based on an Abelian Chern-Simons  theory, that was previously studied in \cite{Aharony:2023zit, Dymarsky:2025agh}. In this limit the genus $\sf g$ partition function can be written explicitly, up to exponentially small corrections.

What makes \VTQFT\, gravity interesting is that in the large central charge limit it exhibits a number of  qualitative features characteristic of semiclassical gravity. This includes the Hawking-Page transition, and the structure of the wormhole amplitude. It also exhibits a toy version of the holographic code. We discuss these properties in detail and outline several lessonsfor the semiclassical gravity.

The paper is organized in a way to help the reader separate technical part from the overall picture. The next section introduces the sum over topologies that defines the bulk theory. It is not technical in nature. Sections \ref{sec:warmup} and \ref{VTQFTgenusreduction} evaluate the bulk sum for an Abelian model first, and then for \VTQFT\, gravity. These sections are the technical heart of the paper. A reader mostly interested in the overall picture may skip these sections and go directly to section \ref{sec:largen}, where the large central charge limit is elucidated. Section \ref{sec:lessons} 
discusses various aspects of \VTQFT\, gravity that make it similar to semiclassical quantum gravity. Going beyond parallels, this section uses similarity between conventional gravity and  TQFT gravity to deduce lessons for the former. 
The final section \ref{sec:summary} contains only a brief summary of the results and a short list of possible future directions.

\section{Sum over topologies and holographic duality} 
\label{sec:sumovert}

The framework of TQFT gravity, that includes the definitions of the bulk theory and the boundary ensemble, and establishes the holographic duality between the two, was developed in \cite{Barbar:2023ncl,Dymarsky:2024frx,Barbar:2025vvf}. We briefly review it below. The relation, and difference, between TQFT gravity and RCFT/CS correspondence is further discussed in \cite{Barbar:2025krh}. 

The TQFT gravity is a model of topological gravity defined by a three-dimensional topological quantum field theory $\T$ summed over topologies. 
To define it 
we first need to introduce some notation.
We denote the Hilbert space of $\T$ placed on a Riemann surface $\Sigma_{\sf g}$ of genus $\sf g$ by $\H_{\sf g}$. We assume $\T$ is unitary and admits topological boundary conditions (TBCs), defined by the Lagrangian algebra objects, that we denote by $\I$. (If $\T$ has non-vanishing chiral central charge, we cancel it by adding the necessary number of $(E_8)_1$ theories.) The path integral of $\T$ on a cylinder $\Sigma_{\sf g}\times [0,1]$ with the TBC $\I$ imposed at one of the  boundaries (symmetry boundary)  defines a state $|\I\rangle_{\sf g} \in \H_{\sf g}$ at the other boundary.  If $\T$ is coupled to a relative 2d theory at the other boundary (which we will call physical boundary), the path integral of $\T$ on the cylinder will evaluate the partition function of some absolute 2d CFT (or TQFT) on $\Sigma_{\sf g}$. That is essentially the SymTFT sandwich construction \cite{Freed:2012bs}. 

Using the sandwich construction we introduce the symmetry factor $|{\rm Aut}(\I)|$ as  the size of the group of invertible symmetries implemented by the Verlined lines on the symmetry boundary. It is a property of $\I$ and does not depend on the choice of the physical boundary. 

Coupling to a relative theory at the physical boundary defines a state ${}_{\sf g}\langle \Omega|\in \H_{\sf g}^*$, where $\Omega$ stands for the period matrix of $\Sigma_{\sf g}$ as well as all other moduli of the 2d theory. Using these notations the CFT partition function defined by the TBC $\I$ can be written as a scalar product between two states, 
\bea
Z_{\I}(\Omega)=\langle \Omega|\I\rangle. 
\eea
We omit the sub-index $\sf g$ of the bra and ket vectors when the corresponding Hilbert space is unambiguous.

Using these notations, the bulk sum over topologies, and the holographic duality between TQFT gravity and the  boundary ensemble can be written as follows,
\bea
\label{sumovertopologies}
\sum_\I {Z_\I(\Omega)\over |{\rm Aut}(\I)|}=\lim_{{g}\rightarrow \infty} {1\over | {\rm MCG}(\Sigma_{g})|} \sum_{\gamma\in {\rm MCG}(\Sigma_{g})} Z_{\T}[C_{{\sf g},g} \cup_\gamma H_{g}].
\eea
On the LHS the sum is over all TBCs of $\T$, which define the boundary ensemble. 
On the RHS, $H_g$ is a handlebody of genus $g$, $\partial H_g=\Sigma_g$, and $C_{{\sf g},g}$ is a compression body -- a geometry with two boundaries,   $\partial_+ C_{{\sf g},g}=\Sigma_{\sf g}, \partial_- C_{{\sf g},g}=\Sigma_{g}$, with the boundary condition ${}_{\sf g}\langle \Omega|$ imposed at the physical boundary $\partial_+ C_{{\sf g},g}=\Sigma_{\sf g}$. The choice of $H_g$ and $C_{{\sf g},g}$ is unimportant. The sum on the RHS is over all generalized Heegaard splittings, 3d geometries with the boundary $\Sigma_{\sf g}$ obtained by gluing $H_g$ to $C_{{\sf g},g}$ after a mapping class group transformation $\gamma$. It is known that when $g\rightarrow \infty$ all connected 3d manifolds with a boundary can be obtained this way. In other words, the bulk sum is a weighted sum over all null-bordisms $M=C_{{\sf g},g} \cup_\gamma H_{g}$ of the boundary Riemann surface $\Sigma_{\sf g}=\partial M$. Relative weights of each $M$ entering the sum are non-trivial and depend on the size of the subgroup of  ${\rm MCG}(\Sigma_{g})$ that leaves $M$ invariant.

We emphasize that the prescription to sum over topologies by averaging over all generalized Heegaard splittings is universal, i.e.~TQFT independent. 

When the boundary is not connected, e.g.~$\Sigma_{{\sf g}_1} \cup \Sigma_{{\sf g}_2}$,  corresponding bulk sum and the holographic equality readily follow from  \eqref{sumovertopologies} by degenerating
\bea
\Sigma_{\sf g}\rightarrow \Sigma_{{\sf g}_1} \cup \Sigma_{{\sf g}_2}, \qquad {\sf g}_1+{\sf g}_2={\sf g}.
\eea

Thinking of the bulk TQFT $\T$ as a SymTFT of some generalized symmetry, the dual boundary ensemble appearing in \eqref{sumovertopologies} can be thought of as the orbifold groupoid of this symmetry \cite{Gaiotto:2020iye}, with different CFTs in the ensemble related by orbifolding.   

We would now like to briefly explain the origin of the bulk sum prescription.
The only  underlying assumption   is as follows. When the genus $g$ is sufficiently large, the only mapping class group invariant states in $\H_g$ are the TBCs states $|\I\rangle_g$ and their linear combinations. This assumption is closely related to an expectation that modular bootstrap at sufficiently large $g$ is powerful enough to allow only well-defined CFTs as its solutions. We also note that for Abelian 3d TQFTs absence of modular-invariant states in $\H_g$ linearly independent from $|\I\rangle_g$ was proven in the context of coding theory \cite{Nebe}.

With this assumption at hand, for sufficiently large $g\rightarrow \infty$ we can decompose the TQFT wavefunction summed over all handlebodies into a linear combination of $|\I\rangle$,
\bea
\label{handlebodies}
\sum_\I p_\I\, |\I\rangle_g = \sum_{\gamma\in {\rm MCG}(\Sigma_{g})} Z_{\rm \T}[\gamma(H_{g})]\, \in\, \H_g,\qquad g\rightarrow \infty. 
\eea
As long as $\T$ is unitary, all coefficients $p_\I$ will be positive. 
This is a formal expression, it disregards an infinite norm of the resulting state.  An expression for smaller genus ${\sf g}< g$ will follow from here  via genus reduction, i.e.~by degenerating the corresponding $\Sigma_g$ into a product of $\Sigma_{\sf g}$ and $g-{\sf g}$ circles. This leads to the following identity,
\bea
\label{nonAmf}
\sum_\I {|\I\rangle_{\sf g}\over |{\rm Aut}(\I)|}=\lim_{{g}\rightarrow \infty} {D^{g-1}\over | {\rm MCG}(\Sigma_{g})|} \sum_{\gamma\in {\rm MCG}(\Sigma_{g})}{}_{g-\sf g}\langle 0|U_\gamma|0\rangle_{g}\, \in\, \H_{\sf g}.
\eea
Here $|0\rangle_{g}\in \H_{g}$ is the ``vacuum'' 
state in the anyon basis, the path integral of $\T$ on a handlebody $H_g$, $\partial H_g=\Sigma_{g}$, without any line insertions. The total quantum dimension $D$ of the theory $\T$ appears because of the normalization ${}_g\langle 0|0\rangle_g=1$. We also introduce $U_\gamma: \H_g \rightarrow \H_g$ to denote the unitary action of the mapping class group. Finally, the scalar product between ${}_{g-\sf g}\langle 0|\in \H_{g-\sf g}^*$ and $U_\gamma|0\rangle_{g} \in \H_g$ assumes an embedding  
\bea
\H_{g_1} \otimes \H_{g_2} \subset \H_{g_1+g_2}. 
\eea
This embedding, as well as the choice of $H_g$ to define $|0\rangle_{g}$ are not unique; different choices are related by the MCG  and hence the resulting expression in \eqref{nonAmf} is not ambiguous. 

The original expression \eqref{sumovertopologies} can be recovered from \eqref{nonAmf} by evaluating its scalar product with ${}_{\sf g}\langle \Omega|$. In other words \eqref{nonAmf} obtained via genus reduction is equivalent to the average over generalized Heegaard splittings \eqref{sumovertopologies}. 

The mapping class group ${\rm MCG}(\Sigma_g)$ is infinite, and in most cases so is its image $U$.
Nevertheless, as long as $\H_g$ is finite-dimensional, both sides of \eqref{nonAmf} are well-defined. This is because  the bulk sum involves the group average, i.e.~it is a projector on the MCG-invariant states in $\H_g$. The overall coefficient ${D^{g-1}\over | {\rm MCG}(\Sigma_{g})|}$ can be thought of as the  wave-function renormalization factor. This overall infinite renormalization factor is crucial for the bulk sum to be well-defined, and to avoid divergences associated with the sum over all bordisms  \cite{Banerjee:2022pmw}. 

We note that even though the initial  sum \eqref{handlebodies} includes only the  handlebodies, the resulting bulk sum \eqref{sumovertopologies} includes all 3d topologies ending on $\Sigma_{\sf g}$. This necessarily follows from the self-consistency of the boundary ensemble under genus reduction. 

A given TQFT $\T$ may yield the same wave-function on different manifolds, $Z_\T[M]\propto Z_\T[M']$, and therefore the bulk sum can be re-organized to go over the equivalence classes of topologies. In some very special cases there is a handlebody in each equivalence class, and therefore 
the whole bulk sum can be written to include only the handlebodies. 
But even in simple cases when $\T$ is an Abelian CS theory, inclusion of non-handlebodies could be necessary. Corresponding geometries can always be represented as handlebodies with the insertions of line operators. To emphasize their conceptual similarity to off-shell topologies in  semiclassical gravity, and the emergence of ``defects'' (line operators) in the bulk, these geometries were called ``singular'' in \cite{Dymarsky:2024frx}. We stress that the bulk sum \eqref{sumovertopologies} includes only smooth geometries $M=C_{{\sf g},g} \cup_\gamma H_{g}$.

So far the sum \eqref{nonAmf} has not been evaluated explicitly barring a few examples when the answer is fixed by symmetry. Those examples include the case of TQFT gravity based on an Abelian CS with a square-free level 
\cite{Aharony:2023zit,Dymarsky:2025agh}. The case of $U(1)$-gravity, that can be obtained as a limit, can also be included in this category, as well as the case of $c=1/2$ Virasoro TQFT  dual to the Ising model \cite{Castro:2011zq,Jian:2019ubz,Romaidis:2023zpx}. When the level of the Abelian CS  is not square-free, the partition function of TQFT gravity can be fixed by matching it to the boundary ensemble \cite{Barbar:2025krh,Angelinos:2025zek}. 

In this work we evaluate the sum (\ref{sumovertopologies},\ref{nonAmf}) explicitly, first in the Abelian case (when the answer is known and fixed by symmetry), and then in the case of \VTQFT\, gravity. We change the overall normalization in \eqref{nonAmf} by multiplying both sides by $D$. The reason is that in the Abelian case $|{\rm Aut}(\I)|=D$ for all $\I$ which makes all weights of the boundary ensemble to be equal to one. In particular, the total number of boundary theories is evaluated by \eqref{nonAmf} with  ${\sf g}=0$, i.e.~by summing over all closed 3d manifolds. More generally, the genus zero calculation evaluates the so-called mass, 
\bea
\label{mass}
M(\T)\equiv\sum_\I {D\over |{\rm Aut}(\I)|},
\eea
the overall normalization of weights of the boundary ensemble. For both the Abelian case and for \VTQFT\, this calculation was recently undertaken in \cite{Dymarsky:2026asf}. We revisit it below.

\section{Warm-up: Abelian Chern-Simons TQFT gravity}
\label{sec:warmup}
Before discussing the case of \VTQFT\, gravity, it is illustrative to first consider a simpler case of  TQFT gravity based on an Abelian Chern-Simons (CS) theory. 
In what follows we will focus on the level $p$ AB theory,  or $\Z_p$ gauge theory,  defined by the action 
\bea\label{AB}
S=\sum_{i=1}^n {p\over 2\pi} \int A_i \wedge dB_i.
\eea
When $p=2$, one copy of this theory describes the ground state of the Toric Code \cite{Kitaev:1997wr}. For that reason we will refer to \eqref{AB} with $p=2$ as the $n$ copies of the Toric Code, or TC${}^n$. 

There are two reasons to start with this Abelian example. First,  the theory above was already discussed in the
context of TQFT gravity in \cite{Aharony:2023zit,Dymarsky:2024frx, Dymarsky:2025agh}. Second, this theory, for $p=2$, is closely related to \VTQFT\, gravity as we will see later. 

Let us briefly recall the structure of the boundary ensemble dual to this theory. 
Topological boundary conditions of an Abelian CS theory are parametrized by the Lagrangian subgroups $\C$, that can be identified with even self-dual codes of a particular type \cite{Barbar:2023ncl, Dymarsky:2026asf}. For that reason we will refer to $\C$ as codes and use this notation instead of $\I$, which is a general label for TBCs in a given topological theory $\T$.
For the theory \eqref{AB} codes $\C$ are defined as follows
\bea
\label{Ccodes}
\sum_i \alpha_i\, \beta_i=0\,\, {\rm mod}\,\, p,\quad {\rm for\, \, any}\,\,\, c=(\alpha,\beta)\in \C\subset (\Z_p\times \Z_p)^n,
\eea 
and 
\bea\label{Ccodes2}
\sum_i \alpha_i \beta'_i+\alpha_i' \beta_i=0\,\, {\rm mod}\,\, p,
\eea
for any $c,c'$ iff $c,c'\in \C$.

Thinking of the theory \eqref{AB} as a SymTFT for an Abelian symmetry ${\mathrm G}=\Z_p^n$,
the boundary ensemble specified by $\langle \Omega|$ consists of boundary CFTs that are related to each other by orbifolding different subgroups of $\Z_p^n \times \Z_p^n$. The resulting ensemble is the orbifold groupoid of \cite{Gaiotto:2020iye}.   
An alternative way to think about the boundary ensemble is as an ensemble of ``code CFTs'' parametrized by $\C$.

The partition function of a CFT $Z_\C$ defined by the TBC $\C$ can be readily written in terms of the enumerator polynomial  (more generally full enumerator) of the code $\C$. This has been discussed in detail in \cite{Dymarsky:2020qom,Angelinos:2022umf,Aharony:2023zit}, when the boundary ensemble consists of a discrete set of Narain theories. We note that the relation between $\Z_\C$ and the enumerator polynomial of $\C$ is more general.  
In particular, one can consider a more general physical boundary by taking any 2d theory with the global symmetry ${\mathrm G}$. For example when $p=2$ one can consider coupling \eqref{AB} to $n$ copies of the 2d Ising CFT \cite{PhysRevLett.114.076402,PhysRevLett.102.220403,Neupert_2016}. The boundary ensemble then consists of various $c=n/2$ CFTs with the chiral algebra ${\rm Vir}_{1/2}^{\otimes n} \times \overline{\rm Vir}_{1/2}^{\otimes n}$ parametrized by $\C$. It was pointed out in \cite{Dymarsky:2024frx} that from the point of view of the TQFT gravity based on \eqref{AB} this boundary ensemble is not fundamental. We will clarify its physical meaning in sections \ref{sec:holographycheck} and \ref{sec:largen} below.

In what follows we evaluate the sum \eqref{nonAmf} for the  theory \eqref{AB} focusing on $p=2$. The result is fixed by symmetry and known, but we will obtain it in a direct way by averaging over the mapping class group ${\rm MCG}(\Sigma_g)$ in the limit of $g\rightarrow \infty$. This will serve as a useful warm-up, preparing the ingredients necessary for the similar calculation in the case of \VTQFT\, gravity.  

The calculation will make heavy use of binary codes, Abelian subgroups of $\Z_2^m$ for some $m$. To help the reader navigate the text we briefly describe them here. First, there are even self-dual codes $\C$ that parametrize the TBCs of \eqref{AB} and hence the CFTs of the boundary ensemble. In the \VTQFT\, case TBCs are no longer parametrized by codes and will be labeled by $\I$ instead. In addition, there are symplectic self-dual codes $\cal S$, which parametrize the wave-function of \eqref{AB} evaluated on different topologies, see \cite{Barbar:2025krh} for a detailed discussion.  These codes are specific for Abelian theories. They can be used to write down the partition function of the Abelian TQFT gravity in a concise way as a Poincar\'e series. 
There are no analogs of $\cal S$ in the \VTQFT\, case, which is a reflection of the fact that the partition function of this theory is more complicated. Finally, in the Abelian case we will see the family of all binary $[n,k]$ codes, i.e.~all possible subgroups $\Z_2^k\subset \Z_2^n$ making an appearance. We denote these codes by $\DD$. These codes have not been discussed in the literature previously, and they are crucial to what follows. In the non-Abelian case of \VTQFT\, gravity these codes will be replaced 
by a different family of binary $[n+\bar n,k]$ codes satisfying additional constraints, that will be denoted by $\D$.

\subsection{Mass (genus zero)}
To begin, we consider a general prime $p$, in which case the image of the mapping class group reduces to $Sp(2g,\Z_p)$, 
and evaluate the mass at genus $g$ defined by

\bea
\label{mainformula}
M_{g,n}=  {{D}^{g}\over |Sp(2g,\Z_p)|} \sum_{\gamma\in Sp(2g,\Z_p)} {}_g\langle 0|U_\gamma|0\rangle_g,\qquad { D}=p^n.
\eea
This quantity was previously evaluated in \cite{Dymarsky:2026asf} by summing over representatives 
of the double-coset $\gamma  \in \Gamma_0 \backslash Sp(2g,\Z_p) / \Gamma_0$,
\bea
\label{massg}
&&M_{g,n}
 =\sum_{h=0}^g \mu(h,g,p)\, D^h=\prod_{i=1}^g  {p^i+p^n\over p^i+1}=p^{gn}\prod_{i=0}^{n-1} {p^i+1\over  p^{i}+p^g},\\
&&\mu(h,g,p)=
{p^{(g-h)(g-h+1)/2} \over \prod_{i=1}^g (p^i+1)}
\prod_{i=1}^h \frac{p^{g-i+1}-1}{p^i-1}. \label{measure}
\eea
In the $g\rightarrow \infty$ limit the mass counts the number of topological boundary conditions, i.e.~the codes $\C$ \cite{Gaiotto:2020iye,Dymarsky:2020qom,Dymarsky:2026asf},
\bea
\label{NC}
N(n)=M=\lim_{g\rightarrow \infty} M_{g,n}=\prod_{i=0}^{n-1}(p^i+1).
\eea
We would like to re-derive this expression in a slightly  different way. 
We introduce the state 
\bea
|W_{g,n}\rangle={{ D}^{g/2}\over |Sp(2g,\Z_p)|} \sum_{\gamma\in Sp(2g,\Z_p)} U_\gamma |0\rangle_g  \in \H_g
\eea
and note that the genus $g$ mass is simply the norm of $|W_{g,n}\rangle$.  Index $n$ is implicit in the definition of $|0\rangle$ which is the vacuum state of the theory \eqref{AB} quantized on $\Sigma_g\times \R$. Up to normalization the state $|W_{g,n}\rangle$ is the averaged full enumerator  of certain symplectic self-dual codes $\cal S$ introduced below. 
It was shown to be equal to the averaged full enumerator of even self-dual codes $\C$ \cite{Dymarsky:2025agh}. This follows from the uniqueness of a state in $\H_g$ invariant under both 
$Sp(2g,\Z_p)$ and the global symmetry group of the CS theory $O(n,n,\Z_p)$.
It is interesting to note that we are interested in calculating the norm of $|W_{g,n}\rangle$ -- a quantity not previously discussed in the coding literature.

The expressions for $|W_{g,n}\rangle$ for $g=1,2$ and arbitrary $n$ are known \cite{Aharony:2023zit,Dymarsky:2025agh}. We instead would like to find the explicit form for arbitrary $g$. It is helpful to switch to a different orthonormal basis \cite{Barbar:2025krh,Dymarsky:2026asf}
\bea
\label{basis}
\kket{(\vec{a},\vec{b})} =\kket{(a_1,b_1)}\dots \kket{(a_n,b_n)} \in \H_g, \qquad a_i,b_i\in \Z_p^{g},
\eea
that is related to the conventional ``anyon'' basis by a Fourier transform. In this basis the group $Sp(2g,\Z_p)$ acts on each $(a_i,b_i)$ as a fundamental vector.   In terms of \eqref{basis} the vacuum state is a ``code'' state (it is  a genus $n$ full enumerator  of code ${\cal S}_0$)
\bea
\label{vacuum}
{D}^{g/2}|0\rangle_g= \sum_{(a_i,b_i)\in {\mathcal S}_0} \kket{(\vec{a},\vec{b})},
\eea
defined by a classical symplectic code ${\mathcal S}_0 \ni (a,0)$ for all $a\in \Z_p^{g}$. When $p$ is square-free the group $Sp(2g,\Z_p)$ maps all such codes to each other. Thus  $|W_{g,n}\rangle$ is the averaged full enumerator over all symplectic codes $\mathcal S$.

To find $|W_{g,n}\rangle$ we would need to classify how the states \eqref{basis}, i.e.~$n$-tuples of vectors in $\Z_p^{2g}$, split into orbits under the action of $Sp(2g,\Z_p).$ For $n=1$ this is particularly simple. There are two orbits, the first includes the zero vector and the second includes all non-zero vectors. 
These orbits are of lengths $1$ and $p^{2g}-1$ correspondingly. If we now write the vacuum state \eqref{vacuum}
\bea
\label{vacuum}
{D}^{g/2}|0\rangle_g=  \kket{(0,0)}+\sum_{a\neq 0}  \kket{(a,0)}, 
\eea
after averaging over $Sp(2g,\Z_p)$ we will readily find
\bea
|W_{g,1}\rangle=\kket{(0,0)}+{p^g-1\over p^{2g}-1}\sum_{(a,b)\neq (0,0)}  \kket{(a,b)}.
\eea
To evaluate its norm we need to sum the squares of all coefficients,
\bea
|W_{g,1}|^2=1+{(p^g-1)^2\over p^{2g}-1}={2p^g\over 1+p^{g}}.
\eea

We would now like to generalize this to arbitrary $n$, focusing on the case of $p=2$. We start with the state 
\bea
\label{initialstate}
\kket{(a_1,0)}\dots \kket{(a_n,0)}
\eea
for some $a_i\in \Z_2^g$ and would like to understand its orbit under the action of $Sp(2g,\Z_2)$. The latter is determined by the linear span of $a_1,\dots,a_n$ (all algebra is understood mod $2$). The span is a $k$-dimensional subspace $\Z_2^k \subset \Z_2^n$ for some $0\leq k \leq n$. Let us introduce a basis in this space $e_1,\dots,e_k\in \Z_2^n$ such that 
\bea
\label{Gdef}
a_i=\sum_{l=1}^k G_i^l\, e_l\, \, {\rm mod}\,\,  2.
\eea
The binary $k\times n$ matrix $G$ can be interpreted as a generating matrix of a binary linear code $[n,k]$ of length $n$ with $k$ generators (here we simply call the linear span of $a_1,\dots,a_n$ written in the basis $e_l$ a code). 
Let us denote this $[n,k]$  code by $\DD$.
Choosing a different basis would yield an equivalent $G$ that generates the same code. Two states of the form \eqref{initialstate} belong to the same orbit under $Sp(2g,\Z_2)$ iff corresponding $\DD$ codes are the same.
This is easy to see by choosing $k$ indices $j_1,\dots,j_k$ such that $a_{j_1},\dots, a_{j_k}$ are linearly independent and then considering their orbit under $Sp(2g,\Z_2)$. 
This logic immediately yields the size of the orbit to be 
\bea
\label{V}
&&V(g,k)=\prod_{i=0}^{k-1}(2^{2g-i}-2^i)=\prod_{i=0}^{k-1} 2^i ({\cal V}(g-i)-1),\quad {\cal V}(g)=2^{2g}-1,
\eea
where ${\cal V}(g)$ is the number of all non-zero vectors in $\Z_2^g$. 

For any given $[n,k]$ code  $\DD$ specified by e.g.~an equivalence class of the generating matrices $G'\sim R\, G$ for any invertible $R\in GL(k,\Z_2)$, there are 
\bea
P(g,k)=\prod_{i=0}^{k-1} (2^g-2^i)
\eea
associated states of the form \eqref{initialstate}. 
The total number of $\DD$ codes, i.e.~subgroups $\Z_2^k\subset \Z_2^n$ is given by the $q$-binomial coefficient
\bea
\label{sigma}
&&\sigma_\DD(n,k)=\left[\begin{array}{c} n \\ k \end{array}\right]_2 = \prod_{i=0}^{k-1} \frac{2^{n-i} - 1}{2^{k-i} - 1},
\eea
when $k>0$ and $\sigma_\DD(n,0)=1$. 
As a consistency check, one can verify this by evaluating the norm of \eqref{vacuum},
\bea
\label{vacuumnorm}
\sum_{k=0}^n \sigma_\DD(n,k) P(g,k)={D}^g=2^{gn}.
\eea
To evaluate the contribution of all states \eqref{initialstate} associated with the same $[n,k]$ code to the genus $g$ mass we can either average over $Sp(2g,\Z_2)$ by choosing representatives in the double-coset $\Gamma_0 \backslash Sp(2g,\Z_2) / \Gamma_0$, or simply by noting that $P(g,k)$ of states associated with the same $[n,k]$ code $\DD$ form an orbit of the size $V(g,k)$,
\bea
\label{consistency1}
\sum_{h=0}^g \mu(h,g,2) P(h,k)={P^2(g,k)\over V(g,k)}.
\eea
The agreement between these two calculations serves as another consistency check. Combined with \eqref{vacuumnorm} this readily gives \eqref{massg}. 

Taking the $g\rightarrow \infty$ limit of \eqref{consistency1} we find the following representation for the mass -- the number of all Lagrangian subgroups $\C$ of the theory \eqref{AB},
\bea
\label{numberN}
N(n)=\sum_{k=0}^n \sigma_{\DD}(n,k)\, 2^{k(k-1)/2}= \prod_{i=0}^{n-1}(2^i+1).
\eea
Physical interpretation of this formula will be given elsewhere \cite{inprogress}.

To summarize, we obtained the previously known result for the mass by introducing and counting $[n,k]$ codes.

\subsection{Torus partition function}
We would now like to extend the discussion above to calculate the torus partition function via genus reduction from some arbitrarily large $g$. What we need to calculate is a state
\bea
|{\rm W}_1\rangle={{D}^g\over |Sp(2g,\Z_2)|}\sum_{\gamma \in Sp(2g,\Z_2)}  {}_{g-1}\langle 0|U_\gamma|0\rangle_g ={D}^{g/2} {}_{g-1}\langle 0|W_{g,n}\rangle\in \H_1.
\eea
Using $Sp(2g,\Z_p)\times O(n,n,\Z_p)$ invariance one can immediately conclude that $|{\rm W}_1\rangle$ is proportional to $|W_{1,n}\rangle$. Indeed, as follows from \eqref{vacuum}, up to normalization the state $|W_{g,n}\rangle$ is a sum of the code states  (full enumerators)
\bea
|W_{g,n}\rangle =\sum_{\gamma \in Sp(2g,\Z_2)/\Gamma_0} |{\cal S}_\gamma\rangle_g,\qquad 
|{\cal S}\rangle_g=
\sum_{a_i,b_i\in {\cal S}} \kket{(\vec{a},\vec{b})},
\eea
where ${\cal S}_\gamma$ is the image of ${\cal S}_0$ under the action of $\gamma$.
One can show that for any $g>g'$ and any two symplectic self-dual codes ${\cal S}_1$ and ${\cal S}_2$ of lengths $g'$ and $g$, respectively, the state
\bea
{}_{g'}\langle {\cal S}_1|{\cal S}_2\rangle_g \in \H_{g-g'},
\eea
is, up to normalization, the code state $|{\cal S}\rangle_{g-g'}$ for some symplectic self-dual code $\cal S$ of length $g-g'$ \cite{Dymarsky:2025agh}. It is then easy to see that $|{\rm W}_1\rangle$ is a sum of the states for all such codes ${\cal S}$ forming an orbit under $Sp(2(g-g'),\Z_2)$. 

This argument is concise but it can not be extended to the more complicated case of  \VTQFT\, gravity. We therefore would like to re-derive this result by evaluating the scalar product between  ${}_{g-1}\langle 0|$ and a particular  state \eqref{initialstate} averaged over $Sp(2g,\Z_2)$ by keeping track of its orbit. Consider a particular $\gamma\in Sp(2g,\Z_2)$. The scalar product between ${}_{g-1}\langle 0|$ and $U_\gamma$-transformed \eqref{initialstate}  will vanish unless for all $i$, $\gamma(a_i,0)^T=(a_i',b_i')^T$ is of the form 
\bea
\label{aftergamma}
a_i'=(\underbrace{*,\dots,*}_{g-1},\alpha_i),\quad b_i'=(\underbrace{0,\dots,0}_{g-1},\beta_i),\quad \alpha_i,\beta_i\in \Z_2.
\eea
Here stars denote arbitrary values.  Provided this condition is satisfied the scalar product will yield the state 
\bea
\label{genus1state}
D^{(g-1)/2}\cdot   {}_{g-1}\langle 0|U_\gamma\kket{(a_1,0)}\dots \kket{(a_n,0)}= \kket{(\alpha_1,\beta_1)}\dots \kket{(\alpha_n,\beta_n)}.
\eea
Since the original $n$-tuple of vectors $(a_i,0)$ are mutually orthogonal with respect to the symplectic product $a\cdot b'-a'\cdot b$, the new vectors will be orthogonal as well, i.e.~ 
\bea
\alpha_i \beta_j-\alpha_j \beta_i =0,
\eea
for any $i,j$.
This implies that if one of the pairs $(\alpha_i,\beta_i)=(1,0)$, all other pairs should also be either $(1,0)$ or $(0,0)$. 

Let us choose $j_1,\dots,j_k$ such that $a_{j_1},\dots,a_{j_k}$ are linearly independent. We further assume that under $U_\gamma$ corresponding  vectors $(a_{j_r},0)$ are mapped to the vectors of the form \eqref{aftergamma} 
with some $\ell$ of them having $(\alpha_j,\beta_j)=(0,0)$ and $k-\ell$ having $(\alpha_j,\beta_j)=(1,0)$.  It is straightforward to calculate the total number of such vectors $(a'_{j_1},b'_{j_1}),\dots, (a'_{j_k},b'_{j_k})$ to be 
\bea
&&\tilde{P}(g,k,\ell)=P(g-1,\ell) P_s(g-1,k-\ell),
\eea
where 
\bea
P_s(g,k)=2^{g} \prod_{i=0}^{k-2}(2^{g}-2^i),
\eea
for $k>0$ and $P_s(g,0)=1$.
With this result at hand we can now calculate the contribution of all vectors \eqref{initialstate} associated with the same binary $[n,k]$ code. For any given choice of $\ell$ values $\alpha_{j_r}=0$ and $k-\ell$ values $\alpha_{j_r}=1$ the coefficient is, c.f.~\eqref{consistency1},
\bea
\label{coefficient}
\lim_{g\rightarrow \infty}{{\tilde P}(g,k,\ell)P(g,k)\over V(g,k)}= 2^{k(k-1)/2-k}.
\eea
There are $2^k$ different choices of $\alpha_j=0,1$ that correspond to $2^k$ codewords of the $[n,k]$ code $\DD$.  Its elements are all possible linear combinations of the rows of $G$. Then corresponding contributions to $|{\rm W}_1\rangle$ combine into  full enumerator  of $\DD$, 
\bea
\label{W1D}
2^{k(k-1)/2-k+n/2}\sum_{(\alpha_1,\dots, \alpha_n)\in {\DD}} 
\kket{(\alpha_1,0)}\dots \kket{(\alpha_n,0)}.
\eea

The full expression for $|{\rm W}_1\rangle$ would also include two additional terms related to \eqref{W1D} by $SL(2,\Z_2)$ permutations, mapping $(\alpha,\beta)=(1,0) \rightarrow (0,1)$ and $(\alpha,\beta)=(1,0) \rightarrow (1,1)$. Care is necessary when dealing with the all-zeros codeword since it is invariant under the action of $SL(2,\Z_2)$.

To make the notations more concise, in what follows we will switch from writing a ket vector in $\H_1$ to writing its wavefunction, assuming the physical boundary $\langle \Omega|$ preserves the permutation symmetry of $n$ copies
\bea
\label{stateOmegaalphabeta}
D^{n/2}\langle \Omega \kket{(\alpha_1,\beta_1)}\dots \kket{(\alpha_n,\beta_n)}=\prod_{i=1}^n \ps_{\alpha_i\beta_i},\qquad D=2.
\eea
In case of the CS theories coupled to Narain CFTs, $\ps_{\alpha\beta}$ are some lattice theta-functions \cite{Barbar:2025krh}. 

Combining all of the above, we find the following torus partition function
\bea
\label{genus1}
{\rm W}_1=\langle \Omega|{\rm W}_1\rangle=\sum_{\DD} 2^{k(k-1)/2-k} \left({W}_{\DD}(\ps_{00},\ps_{10})-{2\over 3}(\ps_{00})^n\right)+{\rm perm},
\eea
where we introduce weight enumerator polynomial of a binary $[n,k]$ code $\DD$,
\bea
W_{\DD}(x,y)=\sum_{c\in {\DD}}x^{n-{w(c)}}y^{w(c)},\qquad w(c)=c_1+\dots+c_n.
\eea
In the expression above permutations stand for two other terms obtained from the first one by the action of $\Gamma_0\backslash SL(2,\Z_2)$, i.e.~by substituting $\ps_{10}\rightarrow \ps_{01}$ and  $\ps_{10}\rightarrow \ps_{11}$.
The extra term $-2/3(\ps_{00})^n$ in \eqref{genus1} is necessary to ensure that when all $\alpha_i=\beta_i=0$ the corresponding state, which is invariant under $SL(2,\Z_2)$, appears with total coefficient one. 

As a consistency check, we note that taking $\ps_{00}=\ps_{10}=1$ will readily give the expression for the mass \eqref{numberN}, $N(n)=M={\rm W}_1(1,1,0,0)$.

The sum above is over all  codes $\DD$ and can be easily evaluated
using the following standard result for the average enumerator polynomial over all $[n,k]$ codes for the given  $n,k$ \cite{macwilliams1977theory},
\bea
{\overline W}_{[n,k]}(x,y) = (1-p_1)x^n + p_1 (x+y)^n,\quad p_1=\frac{2^k - 1}{2^n - 1}.
\eea
Summing over all $0\leq k\leq n$ gives
\bea
\label{g=1AB}
{\rm W}_1=\langle \Omega|{\rm W}_1\rangle={N(n)\over 2^n+2} (\ps_{00}+\ps_{10})^n+(\ps_{10}\rightarrow \ps_{01})+(\ps_{10}\rightarrow \ps_{11}).
\eea

For completeness we note that \eqref{g=1AB}
is precisely the sum over three symplectic codes $\cal S$ of length $2g=2$ 
\bea
{\rm W}_1={N(n)\over 2^n+2}\sum_{\gamma\in SL(2,\Z_2)\backslash \Gamma_0} W_{{\cal S}_\gamma}(\ps_{00},\ps_{10}),\qquad {\cal S}_\gamma=\gamma\, {\cal S}_0,
\eea
which is also equal to averaged enumerator polynomial over all $\C$-codes \cite{Aharony:2023zit,Dymarsky:2025agh},
\bea
\label{averageC}
{\rm W}_1=\sum_{\cal C}W_{\cal C}(\psi_{\alpha\beta}),\qquad \ps_{\alpha\beta}=\sum_{a\in \Z_2} e^{i\pi\, a\, \alpha}\,\psi_{a\beta}. 
\eea

\subsection{Partition function at higher genus }
The calculation of the higher genus partition function is similar but more involved. Focusing on the ${\sf g}=2$  case, we repeat all steps, which are similar to ${\sf g}=1$, except that now $\alpha_j,\beta_j$ are  binary strings of length ${\sf g}=2$. Each initial state \eqref{initialstate}  associated with some $[n,k]$ code $\DD$ will give rise to states
\bea
D^{(g-2)/2}\cdot   {}_{g-2}\langle 0|U_\gamma\kket{(a_1,0)}\dots \kket{(a_n,0)}= \kket{(\alpha_1,0)}\dots \kket{(\alpha_n,0)},
\eea
where each $\alpha_{j}\in \Z_2^2$ takes one of the four values, 
plus the $Sp(4,\Z_2)$ images of the result. For each of $2^{2k}$ possible values of $\alpha_{j_r}$ the corresponding coefficient will be a more complicated version of \eqref{coefficient} and will approach $2^{k(k-1)/2-2k}$ in the $g\rightarrow \infty$ limit. The resulting contribution to ${\rm W}_2$ is then, c.f.~\eqref{genus1},
\bea\nonumber
{\rm W}_2=\langle \Omega|{\rm W}_2\rangle=\sum_{\DD} 2^{k(k-1)/2-2k} \left({W}_{2,\DD}(\ps_{0000},\ps_{1000},\ps_{0100},\ps_{1100})+{16\over 15}(\ps_{0000})^{n}+\right.\\
\left.   -{2\over 3}W_{\DD} (\ps_{0000},\ps_{0100})- {2\over 3}W_{\DD}(\ps_{0000},\ps_{1000})-{2\over 3}W_{\DD}(\ps_{0000},\ps_{1100})
\right)+{\rm perm}, \nonumber
\eea
where permutations stand for another $14$  images 
under $Sp(4,\Z_2)/\Gamma_0$ and extra terms are subtracted to ensure that states with shorter orbits under $Sp(4,\Z_2)$ enter with the correct coefficients. We also introduce the genus 2 enumerator polynomial of a binary code $\DD$,
\bea
W_{2,\DD}(x,y,z,w)=\sum_{c_1,c_2 \in \DD} x^{n-w_1-w_2-w_3} y^{w_1-w_3} z^{w_2-w_3} {w}^{w_3},\\
w_1=w(c_1),\quad w_2=w(c_2),\quad w_3=w(c_1\cdot c_2),
\eea
where
$c_1\cdot c_2$ in the definition of $w_3$ stands for the bit-wise ``and'' operation (bit-wise multiplication). 
Genus $2$ enumerator polynomial averaged over all $[n,k]$ codes for the given $n,k$ has the following form
\bea
\nonumber
\overline{W}_{2,k}(x,y,z,w) &=& (1 - 3p_1 + 2p_2)\, x^{n} + (p_1 - p_2)\Big[(x+y)^n + (x+z)^n + (x+w)^n\Big] +\\
&& p_2\,(x + y + z + w)^n,\qquad p_2 = p_1\frac{(2^{k-1}-1)}{(2^{n-1}-1)}.
\eea
As a consistency check, $\overline{W}_{2,k}(x,y,0,0)=\overline{W}_{k}(x,y)$ and $\overline{W}_{2,k}(x,0,0,w)=\overline{W}_{k}(x,w)$.
Summing over all $0\leq k\leq n$ gives
\bea
\label{g=2AB}
{\rm W}_2=\langle \Omega|{\rm W}_2\rangle={N(n)\over \prod_{i=1}^2 (2^n+2^i)} (\ps_{0000}+\ps_{1000}+\ps_{0100}+\ps_{1100})^n+{\rm perm}.
\eea
It is  equal to averaged enumerator polynomial over all $\C$-codes \cite{Dymarsky:2025agh},
\bea
\label{averageC2}
{\rm W}_2=\sum_{\cal C}W_{2,\cal C}(\psi_{\alpha\beta}),\qquad \ps_{\alpha\beta}=\sum_{a\in \Z_2^2} e^{i\pi\, a\, \alpha}\,\psi_{a\beta}.
\eea
The extension to higher $\sf g$ is straightforward. 

\section{\VTQFT gravity}
\label{VTQFTgenusreduction}
We are now ready to consider the case of interest: TQFT gravity based on multiple copies of a rational Virasoro TQFT with central charge $c=1/2$. 
The terminology of Virasoro TQFT is not quite standard in this context. A more accurate description would be the Reshetikhin-Turaev TQFT associated to the Ising Modular Tensor Category \cite{reshetikhin1991invariants,romaidis2022mapping,Romaidis:2023zpx}, a three-dimensional topological field theory dual to the $c=1/2$ chiral
Virasoro minimal model in the sense of RCFT/CS correspondence \cite{Witten:1988hf,moore1989classical}. For this reason, in what follows we will colloquially refer to this 3d TQFT as Ising theory; 
it should not be confused with the Ising CFT, the full (non-chiral) two-dimensional minimal model $\mathcal{M}(4,3)$.
We consider $n$ chiral and $\bar n$ anti-chiral copies of the Ising theory, with $n-\bar n$ divisible by $16$, supplemented by $(\bar n-n)/16$ copies of $(E_8)_1$ theory to cancel the chiral central charge. 
We will be mostly interested in the case of $\bar n=n$ and will refer to that theory as $n$ copies of the Doubled Ising, or DI${}^n$ for short, although the calculations in this section are general.

The \VTQFT\, gravity is defined as the bulk TQFT -- $n+\bar n$ copies of the Ising theory -- summed over all 3d topologies, as defined in section \ref{sec:sumovert}. Our goal in this section will be to calculate the partition functions of this theory. 

Before we do the bulk calculations, we would like to discuss the boundary ensemble.  
It includes all $c_L=n/2, c_R={\bar n}/2$ CFTs with the  chiral algebra ${\vphantom{\overline{\rm Vir}}{\rm Vir}}_{1/2}^{\otimes n} \times \overline{\rm Vir}_{1/2}^{\otimes \bar n}$
or an extension thereof. For small $n,\bar n$ the space of such CFTs has been studied in the context of framed Vertex Operator Algebras (VOAs) \cite{Dong1994DiscreteSO,dong1998framed,lam2008structure,kitazume2000decomposition,griess2001virasoro,lam2011constructions,lam2012quadratic,lam2015classification,Moriwaki:2021ebe} but it is not thoroughly understood for general  $n,\bar n$. We discuss the case of small $n=\bar n$ in more detail in section \ref{sec:holographycheck}, where we compare the ensemble-averaged torus partition function with its bulk counterpart. 

The CFTs in the ensemble are labeled by $\I$, which parametrize the TBCs of the bulk TQFT. Different $\I$ can give rise to the same 2d CFT represented in different $c=1/2$ Virasoro frames. The sum in \eqref{sumovertopologies} is over all unique TBCs.

There is a well-recognized connection between framed VOAs and codes. While $\I$ themselves are not parametrized by codes, there is a map that assigns each $\I$ a binary code $\D$ that satisfies certain additional properties (includes all-ones codeword and is triply even) \cite{dong1998framed,lam2008structure}. The map is not injective; there are many $\I$ that give rise to the same $\D$. In the discussion below $\D$ will emerge in a very different way, as algebraic objects that parametrize states in the TQFT Hilbert space that are invariant under some subgroup of the mapping class group. 

\subsection{Mass}
The calculation of the mass by summing over all closed 3d manifolds was recently performed in \cite{Dymarsky:2026asf}. We briefly review it here. The starting point is the following basis for $\H_g$ 
\cite{Jian:2019ubz,Alvarez-Gaume:1986rcs}
\bea
\label{basisI}
D^{g/2} \langle \Omega\kket{(a,b)}=t_{ab}\equiv {\vartheta^{1/2} {\genfrac{[}{]}{0pt}{}{b/2}{a/2}}(\Omega|0) / \Phi^{1/2}(\Omega)},\qquad D=2. 
\eea
Here  $a,b\in \Z_2^g$ are binary strings of length $g$ specifying an even spin structure on $\Sigma_g$, 
\bea
a\cdot b=0\,\,{\rm mod}\,\, 2.
\eea
The factor $\Phi(\Omega)$ is independent of $a,b$ and is unimportant for what follows.

The states \eqref{basisI} are canonically normalized and the genus-$g$ vacuum state is given by, c.f.~\eqref{vacuum},
\bea
\label{Ising}
D^{g/2}|0\rangle_g=\sum_{a \in \Z_2^g} \kket{(a,0)},\qquad D=2. 
\eea

The action of the mapping class group on $\H^g$ forms an exact sequence \cite{wright1994reshetikhin, Jian:2019ubz}, 
\bea
\label{sequence}
1\rightarrow U(\Gamma_g)  \rightarrow  U({\rm MCG}(\Sigma_g)) \xrightarrow{\rho} Sp(2g,\Z_2)\rightarrow 1, 
\eea
where $\Gamma_g$ is the subgroup of ${\rm MCG}(\Sigma_g)$ that acts trivially on $\Z_2$-valued homologies $H^1(\Sigma_g,\Z_2)$. 
 Modulo phase factors, the group $Sp(2g,\Z_2)$
acts on the basis \eqref{basisI} by an affine action on the labels, 
\bea
(a,b)^T\rightarrow \gamma\, (a,b)^T+v(\gamma),\qquad \gamma=\left(\begin{array}{cc}A & B\\
C& D\end{array}\right)\in Sp(2g,\Z_2),
\eea
where 
\bea
v(\gamma)=({\rm diag}(B A^T),{\rm diag}(D C^T)),
\eea
and all algebra is understood mod $2$. A simple derivation of this transformation law can be found in \cite{Dijkgraaf:1987jta}.

The kernel of $\rho$, which is a normal subgroup of $U({\rm MCG}(\Sigma_g))$, acts on the basis elements \eqref{basisI} by a pure phase. Since $U(\Gamma_g)$ is a normal subgroup, one can average over the ${\rm MCG}(\Sigma_g)$ in two steps, first over $U(\Gamma_g)$ and then over $Sp(2g,\Z_2)$.  

The discussion so far was for one copy of the Ising TQFT. From now on we consider $n$ chiral and $\bar n$ anti-chiral copies of the Ising theory and start with  the state 
\bea
\label{string}
 \kket{(a_1,0)}  \dots \kket{(a_n,0)} \overline{ \kket{(a_{n+1}, 0)}} \dots  \overline{ \kket{(a_{n+\bar n}, 0)}}, \quad a_i \in \Z_2^g.
\eea
Similarly to the Abelian case, this state defines a binary code as follows. Let us assume there are $k-1\geq 0$ linearly independent differences $a_i-a_j$. 
We can then introduce a basis in the space of differences,  $e_j \in \Z_2^g$, with $j=1,\dots, k-1$, such that, c.f.~\eqref{Gdef},
\bea
\label{ajrep}
a_i=a_1+\sum_{l=1}^{k-1} e_l\, G_i^l, 
\eea
and all algebra is mod $2$. The binary matrix $G$ is $(k-1)\times (n+\bar n)$. It should be extended to be of the size $k\times (n+\bar n)$
by appending a row of ones. The resulting matrix defines an $[n+\bar n, k]$ binary code 
of dimension $k\geq 1$ that we denote $\cal D$. Note that by construction $\cal D$ always includes the all-ones codeword
\bea
c_{\vec{1}}=(\underbrace{1,\dots,1}_n|\underbrace{1,\dots,1}_{\bar n}).
\eea  
It is straightforward to see that choosing a different basis $e_j'$ or choosing another ``reference element'' instead of $a_1$ will yield an equivalent $G'$ generating the same $\cal D$. 

The state \eqref{string} will be invariant under the action of $U(\Gamma_g)$ iff the corresponding code $\cal D$ is triply-even with respect to the modified Hamming weight \cite{Dymarsky:2026asf},
\bea
\label{mhw}
w(c)=w_L(c)-w_R(c),\qquad w_L(c)=\sum_{1}^n c_i,\quad w_R(c)=\sum_{n+1}^{n+\bar n} c_i,\\
c=(c_1,\dots,c_{n}|c_{n+1},\dots,c_{n+\bar n}),
\eea
namely if for all $c\in \cal D$, $w(c)=0\,\,{\rm mod}\, \, 8$. 

To evaluate the genus-$g$ mass we start with the vacuum state 
and average it over $U(\Gamma_g)$,
\bea
\label{emptystate}
{D}^{g/2}|\emptyset\rangle_g={{D}^{g/2}\over |{\rm Ker}(\rho)|} \sum_{\gamma \in {\rm Ker}(\rho)} U_\gamma |0\rangle_g. 
\eea
Only the states \eqref{string} associated with the triply-even $\cal D$-codes survive in the sum. Note, there are $P_s(g,k)$ different states associated with the same $\cal D$. 

Next, we need to average over the action of $Sp(2g,\Z_2)$.
There are two ways to proceed. The first is to sum over the representatives of the double-coset $\Gamma_0 \backslash Sp(2g,\Z_2) / \Gamma_0$. Another is to note that the $P_s(g,k)$ states associated with the same $\cal D$ code, under the affine action of $Sp(2g,\Z_2)$, belong to a single orbit of the size $V_s(g,k)$. 
To evaluate the size of the orbit we note that $Sp(2g,\Z_2)$ acts transitively on all 
\bea
{\cal V}_s(g)=2^{g-1}(2^{g}+1)
\eea
even spin-structures at genus $g$, and the only invariants characterizing the orbit are the symplectic products of the differences $a_i-a_j$. This gives, c.f.~\eqref{V}, 
\bea
&&V_s(g,k)={\cal V}_s(g)\prod_{i=0}^{k-2} 2^i ({\cal V}_s(g-i)-1).
\eea
As a consistency check we verify that both ways of averaging give the same result, 
\bea
\label{consistency1}
\sum_{h=0}^g \mu(h,g,2) P_s(h,k)={P_s^2(g,k)\over V_s(g,k)}.
\eea

After taking the $g\rightarrow \infty$ limit we arrive at the known expression for the mass \cite{Hoehn,Dymarsky:2026asf},
\bea
M=\sum_{k\geq 1} \sigma_{\cal D}(n+\bar n,k)\, 2^{k(k-1)/2+1}.
\eea
Here we introduced $\sigma_\D(n+\bar n,k)$ to denote the total number of all $[n+\bar n,k]$ codes $\cal D$, i.e.~triply-even binary codes  that include the all-ones codeword.  

\subsection{Torus partition function}
The vacuum state $|0\rangle_g$ averaged over the action of $\Gamma_g\subset {\rm MCG}$, see \eqref{emptystate}, is given by the sum over all states \eqref{string} associated with the triply-even codes $\cal D$. Upon a $Sp(2g,\Z_2)$ transformation  such a state specified by $a_1,\dots, a_{n+\bar n}$ would contribute to the torus partition function only if the new labels
\bea
(a_i',b_i')^T=\gamma(a_i,0)^T+v(\gamma),\quad 1\leq i\leq n+\bar n,
\eea
are of the form \eqref{aftergamma}. Note that each $(\alpha_i,\beta_i)$ can take one out of three values $(0,0)$, $(1,0)$, $(0,1)$, but not $(1,1)$, since the latter spin structure is odd. Moreover each expression can include only two out of these three values because 
\bea
(\alpha_i-\alpha_r)(\beta_j-\beta_r)-(\alpha_j-\alpha_r)(\beta_i-\beta_r) =0\,\,{\rm mod}\,\, 2,
\eea
for any $1\leq i,j,r\leq n+\bar n$. Thus the torus partition function will include three terms, first being the polynomial of $t_{00},t_{10}$ (as well as ${\bar t}_{00},{\bar t}_{10}$), second the polynomial of $t_{00},t_{01},{\bar t}_{00},{\bar t}_{01}$  and third of $t_{10},t_{01},{\bar t}_{10},{\bar t}_{01}$.
The first term, which depends on $t_{00},t_{10}$ (and complex conjugate variables), will determine two other terms, which can be obtained by the action of $SL(2,\Z_2)$. 

To evaluate the first term, we can choose $k$ indices $j_r$ such that $a_{j_r}-a_{j_1}$ are linearly independent. Then we need to consider a total of $2^k$ different cases in which some $\ell$ pairs $(\alpha_{j_r},\beta_{j_r})=(0,0)$ and another $k-\ell$ pairs $(\alpha_{j_r},\beta_{j_r})=(1,0)$, for some $\ell$. 
It is straightforward to see that in each case the resulting state is of the form, c.f.~\eqref{genus1state},
\bea
\label{genus1stateV}
D^{(g-1)/2}\cdot   {}_{g-1}\langle 0|U_\gamma\kket{(a_1,0)}\dots \overline{ \kket{(a_{n+\bar n}, 0)}}= \kket{(\alpha_1,0)}\dots \kket{(\alpha_{n+\bar n},0)},
\eea
and corresponds to a particular codeword $(\alpha_1,\dots,\alpha_{n+\bar n})\in {\cal D}$. 
All corresponding coefficients will be the same in the $g\rightarrow \infty$ limit, c.f.~\eqref{coefficient},
\bea
\lim_{g\rightarrow \infty}{P_s(g-1,\ell)P_s(g-1,k-\ell) P_s(g,k)\over V(g,k)}= 2^{k(k-1)/2+1-k}.
\eea
Altogether, this gives the following expression for the torus partition function, c.f.~\eqref{genus1}, 
\bea
{\rm W}_1(t_{00},t_{10},t_{01})=\sum_{\cal D} 2^{k(k-1)/2+1-k}\left(W_{\cal D}(t_{00},t_{10},{\bar t}_{00},{\bar t}_{10})-{t_{00}^n {\overline t}_{00}^{\bar n}+t_{10}^n {\overline t}_{10}^{\bar n}\over 2}\right)+{\rm perm}. \nonumber\\
\label{TPF}
\eea
Here the perm.~denotes two cyclic permutations
\bea
\label{permutationt}
t_{00}\rightarrow t_{10}\rightarrow e^{2\pi i\over 16} t_{01} \rightarrow e^{2\pi i\over 16} t_{00}.
\eea
The terms subtracted from $W_{\cal D}$ are necessary to account for the fact that monomials that only include one variable, e.g.~ $t_{00}^n {\overline t}_{00}^{\bar n}$, and  those that include two, e.g.~$t_{00}^{n-a} {\bar t}_{00}^{\bar n-b}  {\bar t}_{10}^{a} {\bar t}_{10}^b$, understood as vectors in the Hilbert space $\H_1$, such that the order of $t_{00}$ and $t_{10}$ matters, would have different stabilizers in $SL(2,\Z_2)$. We also introduce the modified enumerator polynomial 
\bea
W_{\cal D}(x,y,\bar x,\bar y)=\sum_{c\in {\cal D}}x^{n-{w_L(c)}} {\bar x}^{\bar n -w_R(c)} y^{w_L(c)} {\bar y}^{w_R(c)}.
\eea
Presence of the all-ones codeword implies that $W_{\cal D}$ is symmetric under $x\leftrightarrow y$.

Taking $t_{01}$ to zero would simplify \eqref{TPF},
\bea
{\rm W}_1(t_{00},t_{10},0)=\sum_{\cal D} 2^{k(k-1)/2+1-k} W_{\cal D}(t_{00},t_{10},{\bar t}_{00},{\bar t}_{10}). 
\eea
The full result can be restored if each term is averaged over $SL(2,\Z_2)$ taking into account the size of its stabilizer.  
Further taking $t_{00}=t_{10}=1$ would immediately recover the expression for the mass $M={\rm W}_1(1,1,0)$, 
which is  the coefficient in front of the chiral algebra vacuum character 
\bea
&&{\rm W}_1=M \chi_0^n {\bar \chi}_0^{\bar n}+\dots, \\
&&\chi_0={t_{00}+t_{10}\over 2},\quad \chi_\varepsilon={t_{00}-t_{10}\over 2},\quad \chi_\sigma={t_{01}\over \sqrt{2}}. \label{Isingcharacters}
\eea

\subsection{Partition function at higher genus}
We now briefly discuss the case of ${\sf g}=2$. Starting from the state \eqref{string}, after a transformation specified by some $\gamma$ we find the states of the form \eqref{genus1stateV} as well as their images under $Sp(4,\Z_2)$. 
Thus it is straightforward to see that
\bea
\label{genus2}
\left.{\rm W}_2(t_{ab})\right|_{t_{00b}=0}=\sum_{\cal D} 2^{k(k-1)/2+1-2k}\, W_{2,\cal D}(t_{a00},{\bar t}_{a00}),\qquad a,b\in \Z_2^2,\, b\neq 00. 
\eea
This is a general result: the genus ${\sf g}$ partition function as a function of $t_{a\vec{0}},\, \, a\in Z_2^{\sf g}$, with all $t_{\vec{0}b}$ for $b\neq \vec{0}$ taken to vanish, is given by the sum over genus $\sf g$ enumerator polynomials of the $\cal D$-codes. The definition of $W_{{\sf g},\cal D}$  is straightforward and we only give it for  ${\sf g}=2$,
\bea
\nonumber
&&W_{2,\cal D}(x,y,z,w,{\rm c.c.})=\sum_{c_1,c_2 \in \cal D} x^{n-w_1-w_2+w_3} y^{w_1-w_3} z^{w_2-w_3} {w}^{w_3} {\bar x}^{\bar n-\bar w_1-\bar w_2+\bar w_3} {\bar y}^{\bar w_1-\bar w_3} {\bar z}^{\bar w_2-\bar w_3} {\bar w}^{\bar w_3},\\
&&\qquad w_1=w_L(c_1),\quad w_2=w_L(c_2),\quad w_3=w_L(c_1\cdot c_2),\\ 
&& \qquad {\bar w}_1=w_R(c_1),\quad {\bar w}_2=w_R(c_2),\quad {\bar w}_3=w_R(c_1\cdot c_2).
\eea
We note that $W_{2,\cal D}(x,y,z,w,{\rm c.c.})$ is invariant under all $4!$ permutations of its variables. 

Restoring the dependence on $t_{\vec{0}b}$ for $b\neq \vec{0}$ is also straightforward.  One needs to average \eqref{genus2} or its higher genus generalization over $Sp(2{\sf g},\Z_2)$. The only subtlety is to  take into account 
that terms with different numbers of $t_{a,\vec{0}}$  (understood as vectors in the Hilbert space $\H_{\sf g}$, where the order of variables matters) have different stabilizers within $Sp(2{\sf g},\Z_2)$. Thus the full expression for genus 2 is already quite involved, 
\bea
\label{genus2full}
&& {\rm W}_2(t_{ab})=\sum_{\cal D} 2^{k(k-1)/2+1-2k}\, \Bigg( W_{2,\cal D}(t_{a00},{\bar t}_{a00}) \, + \\
&& {2\over 3}\left(t_{0000}^n {\bar t}_{0000}^{\bar n}+t_{0100}^n {\bar t}_{0100}^{\bar n} +t_{1000}^n {\bar t}_{1000}^{\bar n}+t_{1100}^n {\bar t}_{1100}^{\bar n} \right) -{1\over 2}\sum_{\x,\y\in \{t_{a00}\}}W_{\cal D}(\x,\y,{\bar \x},{\bar\y}) \Bigg) +{\rm perm.}
\nonumber
\eea
The sum on the second line is over the $6$ different ways of choosing $\x\neq \y$ from the set of four variables $t_{a0}$. The permutations denote fourteen additional terms that are obtained from the first one by mapping $t_{a00}$ to $t_{ab}$ using $Sp(4,\Z_2)$.  One can choose, e.g.~the following three generators ${\mathrm g}_i$,
\bea
&&{\mathrm g}_1:\quad t_{0a0b}\rightarrow t_{1a0b} \rightarrow e^{2\pi i \over 16}t_{0a1b}\rightarrow e^{2\pi i \over 16}t_{0a0b},\quad t_{1111}\rightarrow e^{{6\pi i \over 16}}t_{1111},\\ \nonumber
&&{\mathrm g}_2:\quad t_{a0b0}\rightarrow t_{a1b0} \rightarrow e^{2\pi i \over 16}t_{a0b1}\rightarrow e^{2\pi i \over 16}t_{a0b0},\quad t_{1111}\rightarrow e^{{6\pi i \over 16}}t_{1111},\\ \nonumber
&&{\mathrm{g}}_3\colon\;
\left\{
\begin{aligned}
& t_{0000} \to t_{1100} \to e^{\frac{2\pi i}{16}}\,t_{0010} \to e^{\frac{2\pi i}{16}}\,t_{0000},
\quad
t_{0001} \to e^{\frac{6\pi i}{16}}\,t_{0001}, \\[6pt]
& t_{0100} \to e^{\frac{12\pi i}{16}}\,t_{1111} \to e^{\frac{6\pi i}{16}}\,t_{0110}
\to e^{-\frac{6\pi i}{16}}\,t_{0011}
\to e^{-\frac{6\pi i}{16}}\,t_{1000}
\to e^{-\frac{4\pi i}{16}}\,t_{0001}
\to e^{-\frac{4\pi i}{16}}\,t_{0100},
\end{aligned}
\right.
\eea
and the following 15 permutations ${\mathrm g}_1^a {\mathrm g}_2^b {\mathrm g}_3^c$ with: i)  arbitrary $0\leq a,b\leq 2$ and $c=0$, ii) $a=0$,  arbitrary  $0\leq b\le2$, and  $c=1,3$.

\subsection{Comparison with the boundary ensemble}
\label{sec:holographycheck}
Unlike the Abelian case of section \ref{sec:warmup} when the sum over all $\DD$ codes can be explicitly evaluated, there is no known way to enumerate or average over all $\cal D$-codes, binary triply-even codes that include the all-ones codeword. The number of such codes can be evaluated analytically only for small $k$. There is just one such code with $k=1$ and 
\bea
\label{sigma2}
\sigma_\D(n+\bar n,2)={1\over 2}\sum_{w_L=0}^{n} \sum_{w_R=0}^{\bar n} {n!\, {\bar n}!\over (n-w_L)! w_L! ({\bar n}-w_R)! w_R!} \delta_{n+\bar n>w_L+w_R>0}\, \delta_{16\, |\, w_L-w_R}\quad
\eea
for $k=2$.
Writing a similar expression for higher $k$ is possible but it quickly becomes impractical. 


As a warm-up we discuss the chiral case of $n=16$ and $\bar n=0$. In this case all $\cal D$-codes are known \cite{griess2001virasoro}. There are five families with 
\bea
\sigma_\D(16+0,k)=1,\, 6435,\, 2627625,\, 60810750,\, 64864800,
\eea
for $k=1,\dots,5$. Since all codewords, except for the all-ones and all-zeros, have Hamming weight $w=8$, enumerator polynomials are simply 
\bea
W_{\cal D}(x,y)=x^{16}+(2^k-2)x^8 y^8+ y^{16}.
\eea
This gives the following expression for the torus partition function 
\bea
{\rm W_1}=m(t_{00}^{16}+t_{10}^{16}+t_{01}^{16})+\left(M-2m)(t_{00}^8 t_{10}^8-t_{10}^8 t_{01}^8+t_{00}^8 t_{01}^8\right),\\
M=140668954142,\quad m=4643094886.
\eea
The dual ensemble consists of the $(E_8)_1$ theory in many different frames. Taking into account that 
\bea
t_{00}=\sqrt{\theta_3(\tau)\over \eta(\tau)}, \quad t_{10}=\sqrt{\theta_4(\tau)\over \eta(\tau)},\quad 
t_{01}=\sqrt{\theta_2(\tau)\over \eta(\tau)},
\eea
and using standard Jacobi theta-function identity we find ${\rm W_1}=M(t_{00}^{16}+t_{10}^{16}+t_{01}^{16})/2$. Up to an overall coefficient this answer is guaranteed by the uniqueness of $c_L=8, c_R=0$ theory. It can not serve to verify \eqref{TPF} beyond the overall coefficient (mass), which was already established in \cite{Hoehn,Dymarsky:2026asf}. 
The situation with $n=32$ is similar.

The case of $n=48$ is more interesting. In this case \VTQFT\, gravity is dual to an ensemble of 56 different Schellekens theories in different $c=1/2$ Virasoro frames, including the Monster \cite{lam2011constructions,lam2012quadratic,lam2015classification}. Matching the torus partition function in this case would provide a consistency check, albeit a weak one because there is only one parameter not fixed by modular invariance. The necessary ingredients for this check are largely known, but the calculation is involved and we leave this task for the future. 

A non-trivial check that we discuss in detail is provided by $n=\bar n$ theories with small $n$.
When $n=\bar n$, the chiral and anti-chiral Ising theories combine into the so-called Doubled Ising theory; from now on we assume $\bar n=n$.  

The case of $n=1$ DI is special.
In this case there is only one modular invariant, the Ising CFT, and the answer is fixed by modular invariance \cite{Romaidis:2021hnh,Romaidis:2023zpx}. In particular, for $n=\bar n=1$, summing over all bulk topologies as done in this paper, or over only  handlebodies as was done in \cite{Castro:2011zq,Jian:2019ubz,romaidis2022mapping}, with appropriate normalization yields the same result. 
We discuss this calculation in more detail in section \ref{sec:negative}. For completeness, we list the value of $D/{\rm Aut}(\I)=2$ for the Ising CFT, which is the mass of the DI theory.

Modular invariance is not sufficient to fix the partition function when $n>1$. There are multiple topological boundary conditions giving rise to various boundary CFTs. Thus for $n=2$, which corresponds to $c=1$, there are three TBCs. One corresponds to 
``Dirac fermion'' CFT (the compact scalar at radius $R=1$) while two others correspond to its orbifold, the Ising${}^2$ CFT, in two different frames \cite{Ginsparg:1987eb,Dijkgraaf:1987vp}. The coefficients $D/{\rm Aut}(\cal I)$ for these theories are $2$ and $4$ correspondingly, such that the mass of DI${}^2$ is $10$.

A partial classification of all CFTs with the chiral algebra $\chiralalgebra$ for small $c=n/2$ was attempted in \cite{Moriwaki:2021ebe}, but there is no known classification for arbitrary $n$. In the remainder of this section we will discuss a particular family of TBCs $\I$ and hence boundary CFTs that can be obtained as follows. There is a topological interface between the DI theory and the Toric Code -- Abelian CS theory \eqref{AB} with $p=2$ \cite{PhysRevLett.102.220403,PhysRevLett.114.076402}. When $n>1$, $n$ copies of the Doubled Ising (DI${}^n$) have $n!$ different  interfaces with $n$ copies of the Toric Code (TC${}^n$) related by the permutation of copies. Any topological boundary condition of the TC${}^n$, combined with one of the interfaces, gives rise to a topological boundary condition of the DI${}^n$. 
The TBCs of the TC${}^n$ are parametrized by even self-dual codes $\cal C$ defined by (\ref{Ccodes},\ref{Ccodes2}). Different $\cal C$ combined with different interfaces may result in the same TBC of the DI${}^n$. When $n<8$, all TBCs of the DI${}^n$ can be obtained this way. In fact for any $n=\bar n$ this construction gives rise to all TBCs of the DI${}^n$ associated with the $\cal D$-codes that are zero weight, i.e.~obey $w(c)=w_L(c)-w_R(c)=0$ at the level of each codeword $c\in {\cal D}$. 
As long as $n=\bar n<8$, all triply-even codewords automatically have vanishing weight $w(c)=0$. 
Starting from $n\geq 8$ there are additional $\cal D$-codes with codewords $w(c)\neq 0$ and hence additional ways of condensation that should be taken into account \cite{Neupert_2016}.

We would now like to evaluate the ensemble-averaged boundary partition function by summing over those TBCs $\I$ that can be obtained through a particular interface with the TC${}^n$, for all possible $\C$-codes. 
Let us fix a particular pairing of $n$ chiral and $n$ anti-chiral Ising theories, thus fixing an interface between the DI${}^n$ and the TC${}^n$. Each code $\C$, through the interface, would give rise to a TBC $\cal I$, but the map is not injective. Different codes $\C$ that are related by the ``electro-magnetic'' (EM) $\Z_2^n$ symmetry that exchanges individual pairs $\alpha_i \leftrightarrow \beta_i$ of $\C$ would yield the same $\cal I$. Furthermore the symmetry factor $D/{\rm Aut}(\cal I)$ is given by $2^{n-\ell}$ where $2^\ell$ is the size of the EM symmetry group $\Z_2^n$ that leaves the particular code  $\C$ invariant, see the End Matter of \cite{Barbar:2023ncl}.
The resulting CFT partition function $\Z_{\cal I}$ can be expressed in terms of the enumerator polynomial of $\C$ with the appropriate substitution of characters. Thus, for genus one, it is explicitly given by 
\bea
&&Z_{\cal I}(t_{\alpha\beta})=W_{\cal C}(\psi_{\alpha\beta}),\\
&&\psi_{00}\rightarrow {|t_{00}|^2+|t_{10}|^2\over 2},\quad \psi_{10},\psi_{01}\rightarrow {|t_{01}|^2\over 2},\quad \psi_{11}\rightarrow {|t_{00}|^2-|t_{10}|^2\over 2}. \label{substitution}
\eea
The sum over the TBCs $\cal I$ that are associated with the particular interface is given by 
\bea
\label{sum1}
\sum_{\cal I} {4^n\, Z_{\cal I}(t_{\alpha\beta})\over |{\rm Aut}(\cal I)|}= \sum_{[\C]} {2^n\over 2^\ell} W_{{\sf g},\C}(\psi_{\alpha\beta}(t_{\alpha\beta}))=\sum_{\C} W_{{\sf g},\C}(\psi_{\alpha\beta}(t_{\alpha\beta})).
\eea
This result holds for any genus, with $\alpha,\beta\in \Z_2^{\sf g}$. 
The sum in the second expression is over the equivalence classes of $\C$ related by EM symmetry. After rewriting $\psi_{\alpha\beta}$ in terms of $t_{\alpha\beta}$, each code in the equivalence class would yield the same partition function. We can therefore rewrite the expression as a sum over all $\C$-codes. 
This sum can be easily evaluated, see \eqref{averageC} and \eqref{averageC2}.

Our main goal is to match the sum over the TBCs $\cal I$ associated with a given interface \eqref{sum1} with the bulk sum over $\cal D$-codes. 
An even self-dual code $\C$ can be defined by (an equivalence class of) generating matrices, binary $n\times 2n$ matrices of maximal rank,
\bea
(A,B),\qquad A,B\in (\Z_2)^{n\times n},
\eea
satisfying ${\rm diag}(BA^T)=0\,\,{\rm mod}\,\, 2$. A code $\C$ will define the TBC $\cal I$ that in turn defines a code $\D$. The resulting map from $\C$ to $\cal D$  is as follows: $\cal D$ is generated by the $n\times 2n$ matrix (all algebra mod 2),
\bea
(A+B,A+B).
\eea
It is easy to see that the resulting code will only include codewords of zero weight. It is also possible to show that it will include the all-ones codeword. The dimension of the resulting code is $k={\rm rank}(A+B)$. 

In fact any diagonal $\D$-code, i.e.~the one with all codewords of the form $c=(c^+,c^+)$ for some $c^+\in \Z_2^n$, can be obtained from a suitable $\C$.  Any diagonal $\D$-code is a ``doubling'' of some binary  $[n,k]$ code generated by $A+B$. The only condition on the $[n,k]$ code is  that it must include the all-ones codeword. 

For a given $[n,k]$ code containing the all-ones codeword, i.e.~for a given diagonal $\D$-code, there are exactly $2^{k(k-1)/2+1}$ codes $\C$ that map onto it. 
This follows from counting of the constraints on $A,B$ such that $(A,B)$ is of maximal rank, satisfies ${\rm diag}(BA^T)=0$, and $A+B$ is a generator matrix of rank $k$. We skip the proof but as a consistency check we note that the number of all $[n,k]$ codes that include the all-ones codeword is, c.f.~\eqref{sigma},
\bea
&&\sigma_1(n,k)=\left[\begin{array}{c} n-1 \\ k-1 \end{array}\right]_2  = \prod_{i=1}^{k-1} \frac{2^{n-i} - 1}{2^{i} - 1},
\eea
which matches the total number of $\C$-codes \eqref{numberN}
\bea
\sum_{k=1}^n \sigma_1(n,k)\, 2^{k(k-1)/2+1}=N(n)
= \prod_{i=0}^{n-1}(2^i+1).
\eea
In fact there is a stronger statement, at the level of the genus-$\sf g$ enumerators. Namely, the average over all $\C$-codes that correspond to the same $[n,k]$ code (and hence to the same diagonal $\D$-code), after an appropriate substitution of characters $\psi_{\alpha\beta}\rightarrow t_{\alpha\beta}$ is equal to the $\D$-code enumerator. For example, for ${\sf g}=1$, the identity takes the form, 
\bea
\label{hi}
\sum_{\C \in [\D]} W_\C(\psi_{00},0,0,\psi_{11})=2^{k(k-1)/2+1-k}W_{\D}(t_{00},t_{10},0,{\bar t}_{00},{\bar t}_{10},0),
\eea
where the relation between $t_{\alpha\beta}$ and $\psi_{\alpha\beta}$ is given by \eqref{substitution}.
The sum on the LHS is over all $\C$ that correspond to the same diagonal $\D$. 
The full statement with non-vanishing $\psi_{10},\psi_{01},t_{01}$ can be obtained by acting with $Sp(2,\Z_2)$.  The proof will be given elsewhere \cite{inprogress}.
Instead, as a consistency check, one can sum \eqref{hi} over all diagonal $\D$-codes of the given dimension $k$, 
\bea
\label{nk}
&&\sum_{\C} W_\C(\psi_{00},0,0,\psi_{11})= \\
&&2^{k(k-1)/2+1-k}\left({2^{n-1}-2^{k-1}\over 2^{n-1}-1}\left(|t_{00}|^{2n}+|t_{10}|^{2n}\right)+{2^{k-1}-1\over 2^{n-1}-1}\left(|t_{00}|^2+|t_{10}|^2\right)^n\right), \nonumber
\eea
where the sum is over all $\C$ such that ${\rm rank}(A+B)=k$, and the origin of the RHS is further discussed in section \ref{sec:codescount}. Summing this over all $k$ in the range $n\geq k\geq 1$ will readily yield the equality between \eqref{g=1AB} and \eqref{averageC}.

The identity \eqref{hi} is  essentially  the statement that the 
average over the TBCs $\cal I$ that correspond to a particular diagonal code $\D$ is given by that code's enumerator polynomial. 
Writing for genus one explicitly,
\bea\label{hol}
\sum_{\cal I \in [\D]} {4^n\, Z_{\cal I}(t_{\alpha\beta)})\over |{\rm Aut}(\cal I)|}=  2^{k(k-1)/2+1-k} \left(W_{\cal D} -{t_{00}^n {\overline t}_{00}^{\bar n}+t_{10}^n {\overline t}_{10}^{\bar n}\over 2}\right)+{\rm perm}. 
\eea
Note, there is no sum over $\D$ in this formula. 

Up to now we have been discussing the diagonal $\D$-codes. By choosing a different pairing between the $n$ copies of chiral and anti-chiral Ising theories, i.e.~a different interface, one can cover all weight zero $\D$-codes. 
This is based on the observation proven in section \ref{sec:codescount} that all weight zero $\D$-codes  are related to the diagonal ones by permutations $\rho$,
\bea
\rho: (c^+,c^+)\in \D\rightarrow (c^+,\rho(c^+))\in \D',\qquad a\in \Z_2^n.
\eea
From this it readily follows that \eqref{hol} holds for all weight zero $\D$-codes, and for all corresponding TBCs $\cal I$. 

Two different interfaces, related by a relative permutation $\rho$, combined with two different codes $\C$ and $\C'$ (that could be the same), can yield the same TBC $\cal I$. 
That would happen exactly when the corresponding codes  $\D$ and $\D'$ are the same, i.e.~when the corresponding $[n,k]$ code is invariant under the permutation $\rho$ not just as a set but at the level of each individual codeword. This should be taken into account in the sum over all $\I$, which even for $n<8$ is not the same as the naive double sum, one over all $\C$ and another over all $n!$ interfaces.

The identity \eqref{hol}, if established for all $\I$ and all corresponding $\D$-codes, would be sufficient to rigorously  prove the holographic duality between the \VTQFT\, gravity and the boundary ensemble of 2d CFTs for any $n,\bar n$. 
We have only established it for zero weight $\D$ codes, and hence only for $n<8$.  To obtain the statement of holographic duality,  \eqref{hol} or its higher genus generalizations
 should be summed over all $\D$. Writing for genus $g=1$ explicitly, 
\bea
\sum_{\cal I} {4^n\, Z_{\cal I}(\tau)\over |{\rm Aut}(\cal I)|}={\rm W}_1(t_{ab}),
\eea
where ${\rm W}_1$ is defined in \eqref{TPF}.


To summarize, 
we have sketched an analytic argument that the bulk sum precisely matches the ensemble-averaged partition function, as long as all $\D$-codes are zero weight, i.e.~for $n<8$. 
We supplement the argument with a direct computer algebra check for $n \leq 5$. Namely we used the technique outlined above to generate all TBCs  of the DI${}^n$ theory $\I$ starting from the $\C$-codes and interfaces, and made sure to exclude the repetitive $\I$'s from the list. 
We then evaluated corresponding symmetry coefficients $D/{\rm Aut}(\cal I)$ and obtained the ensemble-averaged torus partition function, which was compared with the one obtained directly by summing over $\D$-codes. 

For completeness we list the resulting ${\rm W}_1$, with $t_{01}$ taken to be zero  to save space. Full answer can be obtained using the permutations \eqref{permutationt} after dividing the coefficients of  $|t_{00}|^{2n}$ and $|t_{10}|^{2n}$ by $2$, as in \eqref{TPF}. 
The result is as follows
\begin{alignat}{3}
{\rm W}_1 &= \x + \y,
  &\quad M &= 2, &\quad n &= 1, \nonumber\\
{\rm W}_1 &= 3\x^2 + 4\x\y + 3\y^2,
  &\quad M &= 10, &\quad n &= 2, \nonumber\\
{\rm W}_1 &= 22\x^3 + 45\x^2\y + 45\x\y^2 + 22\y^3,
  &\quad M &= 144, &\quad n &= 3, \nonumber\\
{\rm W}_1 &= 419\x^4 + 1072\x^3\y + 1764\x^2\y^2 + 1072\x\y^3 + 419\y^4,
  &\quad M &= 4746, &\quad n &= 4, \nonumber\\
{\rm W}_1 &= 19906\x^5 + 59325\x^4\y + 130700\x^3\y^2 \nonumber\\
          &\quad + 130700\x^2\y^3 + 59325\x\y^4 + 19906\y^5,
  &\quad M &= 419862, &\quad n &= 5, \nonumber
\end{alignat}
where we introduce $\x=|t_{00}|^2$ and $\y=|t_{10}|^2$. We also list the number of $\cal D$-codes $\sigma_\D(n{+}{\bar n},k)$ necessary for the check, although not the codes themselves.
\begin{table}[h]
\centering
\renewcommand{\arraystretch}{1.3}
\begin{tabular}{c|rrrrr}
\toprule
$n={\bar n} \;\backslash\; k$ & $1$ & $2$ & $3$ & $4$ & $5$ \\
\midrule
$1$ & $1$ & & & & \\
$2$ & $1$ & $2$ & & & \\
$3$ & $1$ & $9$ & $6$ & & \\
$4$ & $1$ & $34$ & $96$ & $24$ & \\
$5$ & $1$ & $125$ & $1250$ & $1200$ & $120$ \\
\bottomrule
\end{tabular}
\caption{Values of $\sigma_\D(n{+}{\bar n},k)$, the number of $[n{+}{\bar n},k]$ triply-even codes that include the all-ones codeword,
for $n = 1, \ldots, 5$ and $k = 1, \ldots, n$.}
\label{tab:dcodes}
\end{table}

\section{Large central charge limit}
\label{sec:largen}
In this section we  focus on the $n=\bar n$ case and consider the limit $n\gg 1$. Our goal is to   evaluate bulk partition function by summing over all $\D$-codes.
To make the presentation more transparent, we first explain the result. 

When $n$ is large, 
almost all of the $\D$-codes are of zero weight; only an exponentially-small fraction of codes have codewords with non-vanishing weight. 
Thus the sum over $\D$'s can be restricted to include only the zero weight codes. Furthermore an exponentially small fraction of such codes are invariant under some permutation $\rho$ acting on the last $n$ bits. Hence, with an exponential precision, the partition function of the bulk theory is given by the naive double sum, first over all diagonal $\D$-codes, and second over all $n!$ permutations that define the interfaces between the DI${}^n$ and  TC${}^n$ theories. 
Since we are interested in the partition function (not a state), the sum over permutations simply gives an overall coefficient $n!$. 

Given the definition of mass  as the overall normalization factor \eqref{mass},  we introduce the normalized ensemble-averaged partition function 
\bea
\avg{Z_{\rm CFT}(\Omega)}\equiv {1\over M}\sum_\I {4^n\, Z_{\cal I}(\Omega)\over |{\rm Aut}(\cal I)|},
\eea
and the same for the bulk partiton function $Z_{\sf bulk}\equiv \langle \Omega|{\rm W}\rangle/M$, such that the holographic identity \eqref{sumovertopologies} takes the form 
\bea
\avg{Z_{\rm CFT}(\Omega)}=Z_{\sf bulk}. 
\eea
As we saw in section \ref{sec:holographycheck}, the sum over all diagonal $\D$-codes can be written as a sum over all $\C$-codes yielding \eqref{g=1AB} or its higher genus generalizations. A more systematic derivation of this result is given below. Thus for $\sf g=1$ we find 
\bea\nonumber
\avg{Z_{\rm CFT}} ={2^n(|\chi_0|^2+|\chi_\varepsilon|^2)^n + \sum_{\ell=0}^1(|\chi_0|^2+|\chi_\varepsilon|^2+2|\chi_\sigma|^2+(-1)^\ell(\chi_0\bar\chi_\varepsilon+\bar\chi_0\chi_\varepsilon))^n\over 2^n+2}.\\
\label{TPFlargec}
\eea
The same expression can be written as a ``Poincar\'e series'' 
\bea\nonumber
\avg{Z_{\rm CFT}}&=&{1\over 1+2^{1-n}} \sum_{\gamma \in \Gamma_0(2)\backslash SL(2,\Z)} \Psi_0(\gamma\, \tau)={\Psi_0(\tau)+\Psi_0(-1/\tau)+\Psi_0(-1/(\tau+1))\over 1+2^{1-n}},
\eea
where $\Psi_0(\tau)=\left(|\chi_0|^2+|\chi_\varepsilon|^2\right)^n$.
The higher genus generalization is straightforward, 
\bea
\avg{Z_{\rm CFT}}={1\over \prod_{h=1}^{\sf g}1+2^{h-\sf g}} \sum_{\gamma \in \Gamma_0(2)\backslash SL(2{\sf g},\Z)} \Psi_0(\gamma\, \Omega),\quad 
\Psi_0(\Omega)=\left(\sum_{a\in \Z_2^{\sf g}} |t_{a\vec{0}}|^2/2^{\sf g}\right)^n. \quad  \label{seedPsdi0}
\eea

The physical picture behind this result is as follows. 
At large central charge the original \VTQFT\, gravity condenses to another theory, TQFT gravity based on the Abelian theory, namely $n$ copies of the Toric Code. All the bulk calculations can be done in the effective bulk theory, coupled to the original boundary state $\langle \Omega|$ through an interface. The interface-based  picture is suggestive of the holographic code, as we discuss in section \ref{sec:lessons}. 

The rest of the section provides a justification for the simplifications outlined above. It is technical and can be skipped at first reading.

\subsection{Triply-even vs doubly-even codes}
We remind the reader that $\D$ codes $[n+\bar n,k]$ of length $n+\bar n$, with $k$ generators are defined as the subgroups of $\Z_2^{n+\bar n}$ that satisfy two additional conditions: i) each codeword $c\in \D \subset \Z_2^{n+\bar n}$ has modified Hamming weight \eqref{mhw} divisible by eight, and ii) the code includes the all-ones codeword $c=(1,\dots,1)$. These codes for $\bar n=0$ were introduced in the context of framed VOAs in \cite{griess2001virasoro}, and generalized for $\bar n\neq 0$ in \cite{Dymarsky:2026asf}. 

The $\D$-codes are closely related, yet very different from the doubly-even $[n+\bar n,k]$ codes, defined in the same way as the $\D$-codes with  one important modification: the modified Hamming weight of each codeword is divisible by four. We note that, by definition, these doubly-even codes include the all-ones codeword. 
The number of these codes is known analytically \cite{Dymarsky:2026asf},
\bea
\label{de}
\sigma_{\sf de}(n+\bar n,k)= \prod_{i=1}^{k-1} {(2^{(n+\bar n)/2-i-1}+1)(2^{(n+\bar n)/2-i}-1)\over (2^i-1)}.
\eea
Since any triply-even code is also doubly-even, \eqref{de} provides an upper bound on the number of $\D$-codes. But unlike the doubly-even case, the exact number of $\D$-codes of a given length  is not known. The ``chiral'' $\D$-codes with $\bar n=0$ of moderate lengths were studied extensively \cite{betsumiya2012triply}.

There is a simple reason for the difference between doubly-even and triply-even codes: there is a group action on $\Z_2^{n+\bar n}$ that preserves the double evenness of the codewords, but the group becomes essentially trivial if the Hamming weight mod $8$ is preserved. To further illustrate the difference between doubly-even and triply-even codes, we note that for $\bar n=0$, whenever $n$ is divisible by $8$, there are doubly-even self-dual codes, the so-called Type II binary codes that emerge in physics as the Lagrangian subgroups of $U(1)_2^n$ CS theory. Their number follows from \eqref{de} and is given by $N(n-1)$, where $N(n)$ is defined in \eqref{numberN}.
Yet, for $\bar n=0$, there are no self-dual triply-even codes for any $n$, i.e.~the number of generators $k$ is always strictly smaller than $n/2$. 
A way to see this is to note that the enumerator polynomial $W(x,y)$ of a doubly-even self-dual code must be invariant under $T$ and $S$ transformations 
\bea
&&T: x\rightarrow x,\quad y\rightarrow i\,y, \\
&&S: x\rightarrow {x+y\over \sqrt{2}},\quad y\rightarrow {x-y\over \sqrt{2}}.
\eea
These are modular transformations of the two $U(1)_2$ characters, and the full group (modulo its center which acts trivially on $W(x,y)$) is the group of modular transformations $\Gamma(2)\backslash SL(2,\Z)$ of the $U(1)_2$ CS theory. As established by the Gleason's theorem, the ring of polynomials invariant under $T,S$ is freely generated by two elements \cite{Nebe}. We also note that the group generated by $T,S$ is finite, it is the Clifford group of one spin. This too can be understood from the $U(1)_2^n$ CS theory. The Wilson line operators in the $U(1)_2$ theory quantized on a torus are the Pauli operators (Clifford gates) and the modular group therefore must be a subgroup of the Clifford group. 

Similar considerations applied to the triply-even case lead to dramatically  different conclusions. Self-dual triply even codes must be invariant under $S$ (which is MacWilliams identity at the enumerator polynomial level, or the Hadamard gate in terms of the quantum spin) and the so-called T-gate
\bea
\tilde{T}: x\rightarrow x,\quad y\rightarrow e^{i\pi/2}\,y.
\eea
It is well-known that inclusion of the T-gate leads to a universal set of quantum gates, and the corresponding group generated by $\tilde{T},S$ is infinite. Accordingly there is no $W(x,y)$ invariant under both $S$ and $\tilde T$, which means self-dual triply even $[n,n/2]$ codes do not exist.  It is interesting to note that the appearance of the T-gate leads to an interesting connection with the quantum error-correction literature 
 \cite{Bravyi:2015suv,Haah:2018uxe, Rengaswamy:2020fyi,Jain:2024zdq}.

\subsection{Counting $[n+n,k]$ $\D$-codes}
\label{sec:codescount}
In the reminder of this section we focus on the $\bar n=n$ case and discuss $[n+n,k]$ $\D$-codes. 

In addition to triply-even $\D$ codes and the doubly-even codes introduced above, there is another related family of zero weight codes. It was already introduced in section \ref{sec:holographycheck} but we repeat the definition here. A zero weight code is a $k$-dimensional subgroups of $\Z_2^{n+n}$ satisfying: i) all codewords have vanishing modified Hamming weight \eqref{mhw}, and ii) 
all codes include the all-ones codeword. There is a natural inclusion: 
\bea
{\rm zero\,\, weight\,\, codes}\, \subset\, \D-{\rm codes}\, \subset\, {\rm doubly\,\, even\,\, codes}.
\eea
The number of the zero weight codes can be estimated as follows. First, we note that using an appropriate permutation of the last $n$ letters each zero weight code can be brought to the diagonal form
\bea
c=(c^+,c^+)\in \D\subset  \Z_2^{n+n},\quad c^+\in \Z_2^n.
\eea
Indeed, since all codewords are of zero weight, for any $c=(c^+,c^-)\in \D$, we have $w(c^+)=w(c^-)$, where $w$ is the standard Hamming weight on $\Z_2^n$. By considering $k$ generators $c_i=(c^+_i,c^-_i)$, a zero-weight code defines a linear Hamming weight-preserving map from $\Z_2^n$ to $\Z_2^n$, where 
a linear combination of $c^+_i$ is mapped to a linear combination of $c^-_i$. According to MacWilliams extension theorem this map is a restriction of a permutation $\rho: \Z_2^n \rightarrow \Z_2^n$. 

A zero weight $\D$-code with all codewords of the form $c=(c^+,c^+)\in \D$ will be called diagonal. 
As we already mentioned in section \ref{sec:holographycheck} the  diagonal $\D$ codes are in one to one correspondence  with the binary $[n,k]$ codes that include the all-ones codeword. Their number is given by 
\bea
\sigma_1(n+n,k)=\left[\begin{array}{c} n-1 \\ k-1 \end{array}\right]_2=\prod_{i=1}^{k-1 }{2^{n-i}-1\over 2^i-1}.
\eea
The number of zero-weight codes can be obtained by counting the orbit size of each diagonal $\D$-code under the permutation group acting on $n$ last bits. This count is simplest for $k=n$, i.e.~for the self-dual codes (which exist in the symmetric case $\bar n=n$, unlike the chiral case $\bar n=0$). Indeed in this case there is a unique diagonal $\D$-code with the $n\times 2n$ generator matrix 
\bea
(I|I).
\eea
All $n!$ permutations yield non-trivial self-dual codes with the same enumerator polynomial $W_\D(t_{00},t_{10},{\bar t}_{00},{\bar t}_{10})=(|t_{00}|^2+|t_{10}|^2)^n$, the only polynomial invariant under 
\bea
&&\tilde{T}: t_{00}\rightarrow t_{00},\quad t_{10}\rightarrow e^{i\pi/2}\,t_{10}, \\
&&S: t_{00}\rightarrow {t_{00}+t_{10}\over \sqrt{2}},\quad t_{10}\rightarrow {t_{00}-t_{10}\over \sqrt{2}}.
\eea
Thus $\sigma_{\sf zw}(n+n,n)=n!$.

The analysis for $k=n-1$ is a bit more involved. A diagonal $[n+n,k]$ code is defined by its projection on the first $n$ letters, i.e.~by a $[n,k]$ binary code that includes the all-ones codeword. For $k=n-1$ all such codes are specified by their parity check vector $h\in \Z_2^n$, the vector orthogonal to all codewords (i.e.~the vector generating the dual $[n,1]$ code). The parity check vector is arbitrary except that it must have an even Hamming weight for the dual code to include the all-ones codeword. There are $\sigma_1(n,n-1)=2^{n-1}-1$ such codes.   
When the Hamming weight $w(h)=2$, the permutation of the two bits in the support of $h$ is a symmetry of the $[n,n-1]$
code at the level of each codeword. Hence the orbit of the corresponding diagonal $\D$-code under the permutation group $\rho:\Z_2^n \rightarrow \Z_2^n$ is of the size $n!/2$. When the Hamming weight $w(h)\geq 4$, any permutation that leaves the support of $h$ invariant is a symmetry of the $[n,k]$ code as a set, but there are always codewords not invariant under any particular $\rho$. Therefore for $w(h)\geq 4$ the orbit of the diagonal $\D$-code includes $n!$ different zero weight $\D$-codes.
This gives the following expression for the number of $[n+n,n-1]$ zero weight codes, 
\bea
\sigma_{\sf zw}(n+n,n-1)=\sum_{l=1}^{[n/2]}{(n!)^2\over (2l)! (n-2l)!} \left(1-{1\over 2}\delta_{l,1}\right)= \!\left(2^{n-1} - 1 - \frac{1}{2}\binom{n}{2}\right).
\eea
The main  message here is that the $[n,k]$ codes invariant under some of the permutations at the level of all codewords amount to an exponentially small fraction among all $[n,k]$ codes. Hence, at leading order $\sigma_{\sf zw}(n+n,n-1)\approx n! \,\sigma_1(n,n-1)$, 
and the same logic applies to all smaller $k$, 
\bea
\sigma_{\sf zw}(n+n,k)\approx n!\, \sigma_1(n,k),
\eea
while corrections are exponentially suppressed. 
The same result applies to the averaged enumerator polynomial: $\overline{W}_\D(t_{00},t_{10})$ averaged over all $[n+n,k]$ zero weight codes is approximately given 
by the averaged enumerator polynomial of all $[n,k]$ codes that include the all-ones codeword
\bea
&&\overline{W}_\D(t_{00},t_{10})\approx \overline{W}_{[n,k]}(|t_{00}|^2,|t_{10}|^2),\\
&&\overline{W}_{[n,k]}(x,y)=\frac{2^{n-1}-2^{k-1}}{2^{n-1}-1}(x^n+y^n)
        + \frac{2^{k-1}-1}{2^{n-1}-1}(x+y)^n.
\eea
If we now assume that the zero weight codes dominate the space of all $\D$-codes, an assumption that will be justified below, we readily find the average over all $\D$-codes to be
\bea
{\rm W}_1&=&\sum_{\D} 2^{k(k-1)/2+1-k} \left(W_\D(t_{00},t_{10})-{|t_{00}|^{2n}+|t_{10}|^{2n}\over 2}\right)+{\rm perm.}\approx \\ 
&&\sum_{k=1}^n 2^{k(k-1)/2+1} n!\, \sigma_1(n,k) \left(\overline{W}_{[n,k]}(|t_{00}|^2,|t_{10}|^2)-{|t_{00}|^{2n}+|t_{10}|^{2n}\over 2}\right)+{\rm perm.}= \nonumber \\
&&{n!\, N(n)\over 2^n+2}\left(|t_{00}|^2+|t_{10}|^2\right)^n + {\rm perm.} \nonumber
\eea

Finally we want to estimate the number of $\D$-codes that are not zero weight. We start with the doubly-even codes that are fully under control. For $n<3$, all $[n+n,k]$ doubly-even codes are automatically zero weight. Starting from $n\geq 4$ there could be codewords with $w_L(c)-w_R(c)=4m$. For simplicity we count the number of doubly-even codes that include just one such codeword $c_0=(c_0^+,c_0^-)$ with $m=1$ and $w(c^+_0)=4, w_R(c^-_0)=0$. All such codes can be obtained as follows. The codeword $c_0$ is combined with a diagonal  doubly-even code based on some $[n,k-1]$ code with the all-ones codeword to define a $[n+n,k]$ doubly-even code. All other codes can be obtained by a permutation of the last $n$  bits. The only necessary condition is that the projection  of the $[n,k-1]$ code on the support of $c_0^+$ should be even. 

The number of all such $[n,k-1]$ codes is easy to find out. Even projection on the support of $c_0^+$ means the $[n,k-1]$ code is orthogonal to $c_0^+$, which is one linear constraint. The $[n,k-1]$ code must includes the all-ones codeword, which can be chosen as one of the generators and imposes another linear constraint. Hence the total number of such $[n,k-1]$ codes is 
\bea
\left[\begin{array}{c} n-2 \\ k-2 \end{array}\right]_2.
\eea
This implies that $k$ ranges between $2\leq k\leq n$, and in particular the number of self-dual doubly-even non-diagonal codes that include $c_0$ will be 
\bea
{n!\over 4! (n-4)!}n!  \left[\begin{array}{c} n-2 \\ n-2 \end{array}\right]_2\sim  n!\, n^4.
\eea
Here the first factor $\binom{n}{4}$ reflects all possible choices of $c_0$, while $n!$ stands for all possible permutations of the last $n$ bits. We note this number is already by a factor of $n^4$ larger than the number of the zero weight self-dual codes, which is $n!$. The overall picture is that as $n$ grows, the total number of non-zero weight doubly-even codes dominates  \eqref{de} and vastly surpasses the number of zero weight codes, 
\bea
\sigma_{\sf de}(n+n,k)\approx 2^{(2n-3k/2-1)(k-1)+O(1)} \gg \sigma_{\sf zw}(n+n,k)\approx n! \, 2^{(n-k)(k-1)+O(1)}.
\eea

A similar calculation for the triply-even $\D$-codes reveals a qualitatively very different picture. For $n<8$, all $[n+n,k]$ $\D$-codes are zero weight, but starting from $n\geq 8$ there could be codewords $c_0=(c_0^+,c_0^-)$ with $w_L(c)-w_R(c)=8m$. Let us estimate the number of such codes with $m=1$ and $w(c^+_0)=8, w_R(c^-_0)=0$. It is controlled by the number of $[n,k-1]$ codes that include the all-ones vector and have the projector on the support of $c_0^+$ which is doubly-even. This condition is qualitatively different from the condition of evenness encountered above. In particular, $k-1$ is ranging between $1\leq k-1\leq n-4$. To see the origin of this upper bound, we assume that the projection of $[n,k-1]$ code on the support of $c_0^+$ is maximal possible, which is up to permutations unique -- it is the self-dual $[8,4,4]$ Hamming code. Thus all such $[n,k-1]$ codes can be defined as $(k-1)$-dimensional spaces orthogonal to $[8,4,4]$ Hamming code which is extended by zeros to be of length $n$. This imposes $4$ linear constraints. Amended with another linear constraint that the $[n,k-1]$ code includes the all-ones codeword, it yields the following count for the number of $[n,k-1]$ codes, 
\bea
\left[\begin{array}{c} n-5 \\ k-2 \end{array}\right]_2.
\eea
From here we readily see that $k$ can not be larger than $n-3$, i.e.~all self-dual $[n+n,n]$ and $[n+n,n-1]$ $\D$-codes must be zero-weight.
(That all self-dual $\D$-codes must be zero weight readily follows from the form of the unique enumerator polynomial invariant under $\tilde{T},S$.)

We now estimate the number of non-zero weight 
$[n+n,n-3]$ $\D$-codes that include only one non-zero weight vector $c_0$. It is given by
\bea
{n!\over 8!(n-8)!} n! N(2) \left[\begin{array}{c} n-5 \\ n-5 \end{array}\right]_2 \sim n!\, n^8.  
\eea
Here $\binom{n}{8}$ reflects all possible choices of $c_0$, $n!$ stands for possible permutations of the last $n$ bits and $N(2)=30$ stands for $30$ possible permutations of the Hamming code. 
Comparing with the number of  zero weight $[n+n,n-3]$ codes $\sigma_{\sf zw}(n+n,n-3)\sim n!\, 2^{3(n-4)}$, we find the latter to be exponentially larger. Consideration more general non zero weight $\D$-codes confirm the overall picture:  the number of non zero weight codes is exponentially suppressed. 
 

\section{Parallels with and lessons for quantum gravity}
\label{sec:lessons}
The model discussed in this paper, \VTQFT\, gravity, in the limit of large central charge exhibits many qualitative features similar to those expected in semiclassical quantum gravity. We discuss them below to outline  parallels and draw lessons.

\subsection{``Off-shell'' topologies cure negative density of states}
\label{sec:negative}
The sum over classical saddle contributions to the 3d gravity torus partition function famously suffers from the negative density of states \cite{Maloney:2007ud,Keller:2014xba,Benjamin:2019stq}. It is a well-established expectation that the contributions of the off-shell topologies to the bulk path integral, e.g.~conical defects, will cure these negativities \cite{Maxfield:2020ale,Benjamin:2020mfz,Alday:2019vdr,DiUbaldo:2023hkc}. The   \VTQFT\, gravity provides an explicit example realizing this scenario.

In the context of TQFT gravity, there is no notion of the off and on-shell topologies. Instead we compare the sum over handlebodies with the full sum \eqref{sumovertopologies}. Thus, for genus $\sf g$ we introduce 
\bea
\label{naivebs}
Z_{\sf hb}(\Omega)= {1\over | {\rm MCG}(\Sigma_{\sf g})|} \sum_{\gamma\in {\rm MCG}(\Sigma_{\sf g})} {}_{\sf g}\langle \Omega|U_\gamma|0\rangle_{\sf g},
\eea
which should be compared with the full answer evaluated in section \ref{VTQFTgenusreduction}.

Uniqueness of the modular invariant for $n=\bar n=1$, namely the Ising CFT partition function,  implies that 
\bea
Z_{\sf hb}(\Omega_g)={Z_{\sf Ising}(\Omega_g) \over 2^{\sf g-1}(2^{\sf g}+1)},
\eea
where the normalization factor comes from the norm of $Z_{\sf Ising}$ understood as a state in $\H_{\sf g}$. 
It is simply the number of even spin structures at genus $\sf g$. This is in agreement with the explicit calculations of \cite{Castro:2011zq,Jian:2019ubz} for ${\sf g}=1,2$.

When  $n=\bar n>2$ the naive bulk sum \eqref{naive} can not be interpreted as an ensemble average because of the negative density of states. Comparing with the calculation of section \ref{VTQFTgenusreduction}, we can immediately conclude that starting with \eqref{naivebs}, only the states \eqref{string} associated with the $\cal D$-codes of dimension $k\leq {\sf g}$ may contribute to the sum. An even more crucial difference is in the coefficient with which each term contributes. 
Were we to calculate the torus partition function starting from \eqref{naive} by genus reduction, each codeword of $\cal D$ would appear with the coefficient \eqref{coefficient} but without taking the $g\rightarrow \infty$ limit. To see how much that affects the result, we calculate the torus partition function starting from \eqref{naivebs} with $\sf g=1$. This is a straightforward extension of the calculation performed in \cite{Castro:2011zq},
\bea
\nonumber
Z_{\sf hb}&=&\sum_{w_L=0}^{n} \sum_{w_R=0}^{\bar n}  {n!\, {\bar n}!\, \delta_{16\, |\, w_L-w_R}\over (n-w_L)! w_L! ({\bar n}-w_R)! w_R!}t_{00}^{n-w_L} t_{10}^{w_L} {\bar t}_{00}^{\bar n-w_R} {\bar t}_{10}^{w_R} +{\rm perm.}
\eea
Comparing with \eqref{sigma2} we indeed find that the result can be rewritten as a sum over $k=1$ and $k=2$ $\cal D$-codes but with unnatural coefficients. 

If  written in terms of the Ising characters, this expression develops negative density of states,
\bea
\nonumber
Z_{\sf hb}=\sum_{w_L=0}^{n} \sum_{w_R=0}^{\bar n} {n!\, {\bar n}!\, \delta_{16\, |\, w_L-w_R}\over (n-w_L)! w_L! ({\bar n}-w_R)! w_R!}\, \, \left(\chi_0^n {\bar \chi}_0^{\bar n} +{(n-2w_L)^2-n\over 2}\chi_0^{n-2}\chi_\varepsilon^2 {\bar \chi}_0^{\bar n}+  \dots \right).
\eea
The main contribution to the sum comes from $w_L\approx n/2$ when the coefficient in front of the second term is negative. This should be contrasted with the exact result \eqref{TPF} which is manifestly positive, if written in terms of the Ising characters \eqref{Isingcharacters},
\bea
Z_{\sf bulk} &\propto \sum_{\cal D} 2^{k(k-1)/2+1-k} \biggl(W_{\cal D}\left(\chi_0+\chi_\varepsilon,\chi_0-\chi_\varepsilon\right) + \sum_{l=0}^1 W_{\cal D}\left(\chi_0+(-1)^l\chi_\varepsilon,\sqrt{2}\chi_\sigma\right) \nonumber\\
&\hspace{6em} -\sum_{l=0}^1 |\chi_0+(-1)^l\chi_\varepsilon|^{2n}-|\sqrt{2}\chi_\sigma|^{2n} \biggr). 
\eea
To see that we note that the first term inside the parenthesis is the enumerator polynomial of the dual code $\D^\perp$, as follows from the 
MacWilliams identity, 
\bea
W_{\cal D}\left(\chi_0+\chi_\varepsilon,\chi_0-\chi_\varepsilon\right)=2^k W_{\cal D^\perp}\left(\chi_0,\chi_\varepsilon\right).
\eea
(We suppressed complex conjugate variables in the argument for simplicity.)
Hence all the coefficients in the decomposition of $Z_{\sf bulk}$ in terms of the Ising characters are manifestly positive. The appearance of the enumerator of $\D^\perp$ is not accidental. 
It has been shown that this is precisely the binary code that keeps track of the terms in the CFT torus partition function after all $\chi_\sigma$ are taken to vanish \cite{dong1998framed}.

As a final comment, we note that when $n\gg 1$, the effect of the full sum \eqref{sumovertopologies} can be easily quantified as follows. The naive bulk sum over the handlebodies \eqref{naivebs} is the Poincar\'e series (the sum over modular transformations) of the seed 
\bea
\langle \Omega|0\rangle_{\sf g} =\left|\sum_{a\in \Z_2^{\sf g}} t_{a\vec{0}}\right|^{2n}/2^{n\sf g},
\eea
while the correct seed that takes into account all possible topologies is \eqref{seedPsdi0},
\bea
\Psi_0 \propto \left(\sum_{a\in \Z_2^{\sf g}} |t_{a\vec{0}}|^2\right)^n/2^{n\sf g}.
\eea

\subsection{Prospects of a semiclassical description}
One interesting feature of the  torus partition function \eqref{TPFlargec} in the large central charge limit is its structure consistent with three saddle point contributions and the presence of the Hawking-Page transition. The origin of this structure is the Abelian theory \eqref{AB} with $p=2$ that combines all topologies with the torus boundary into three equivalence classes (parametrized by three symplectic codes $\cal S$ of length 2). It is tempting to ask if perhaps \VTQFT\, gravity in the large central charge limit can be described semiclassically, by some weakly coupled bulk theory. Even if such a description is possible, it will not be simple. 

First, all 2d theories from the boundary ensemble are not sparse in the conventional $c=n/2$ Virasoro sense of \cite{Hartman:2014oaa}. This is because the Ising vacuum character $|\chi_0|^{2n}$ contains many  $c=n/2$ Virasoro primaries, as evident from the corrections to the Cardy formula. Starting from the $S$-invariant expression 
\bea
Z=|\chi_0(\tau)|^{2n}+|\chi_0(-1/\tau)|^{2n},
\eea
while ignoring any possible corrections, one finds for the free energy at leading $1/c$ order
\bea
{1\over c}\ln Z\approx {\pi^2\over 3\beta} +4 e^{-8\pi^2/\beta}+\dots ,\qquad \tau=i{\beta\over 2\pi}, \quad \beta\leq  2\pi. 
\eea
The corrections are exponentially suppressed for small $\beta$ but not in $1/c$. 
From this we conclude that 
the hypothetical semiclassical bulk description, in addition to gravity, should include $O(c)$ additional tensor gauge fields, to account for $n$ conserved spin-2 currents at the boundary.

It is interesting to note that the full result  \eqref{TPFlargec} exhibits different  free energy asymptotics
\bea
{1\over c}\ln \avg{Z_{\sf CFT}}\approx {\pi^2\over 3\beta} +2 e^{-4\pi^2/\beta}+\dots ,\
\eea
suggesting the dual boundary ensemble is not sparse even in terms of the Ising primaries. At first this conclusion is surprising, given that in the large $c$ limit the boundary ensemble is parametrized by $\C$-codes. The standard expectation in the context of codes is that a random code would have large minimal weight (Hamming distance), proportional to  code's length. This  is the so-called Gilbert–Varshamov bound. In the context of code-based CFTs this means the ensemble-averaged spectral gap (defined in terms of the appropriate characters/primaries) will be linear in central charge \cite{Henriksson:2022dml,Angelinos:2022umf}, suggesting sparseness. What happens in our case is that boundary CFTs, on average, have a large spectral gap in terms of the modified primaries/characters $\psi_{ab}$ \eqref{substitution}, but not in terms of the Ising primaries/characters $\chi_0$ and $\chi_\epsilon$. In particular the ``vacuum character'' $\psi_{00}^n$ already contains many light states, which are primaries of $\chiralalgebra$.

\subsection{Wormhole amplitude}
Another interesting quantity to calculate is the double-sided wormhole amplitude, the full connected part of the torus partition function
\bea
\label{wh}
{\Wh}(\tau_1,\tau_2)=\avg{Z_{\sf CFT}(\tau_1)Z_{\sf CFT}(\tau_2)}-\avg{Z_{\sf CFT}(\tau_1)}\avg{Z_{\sf CFT}(\tau_2)}.
\eea
In the large central charge limit, at leading order,  this calculation can be done in the Abelian TQFT gravity, by evaluating the connected part of the genus $2$ enumerator polynomial of the $p=2$ $\C$-codes. 
This calculation was carried out in \cite{Dymarsky:2020pzc},\footnote{The ensemble of $\C$ codes discussed in \cite{Dymarsky:2020pzc} was slightly different from the one relevant here; this difference is unimportant in the large $n$ limit.} which established that for $n\gg 1$ the ensemble of $\C$ codes is self-averaging. In other words the connected part \eqref{wh} is exponentially suppressed with the central charge $\Wh\sim e^{-O(c)}$. This behavior is expected in  semiclassical gravity, but is more general and only reflects the fact that the boundary ensemble of CFTs is self-averaging.

It is illustrative to evaluate $\Wh$ explicitly. We do that in the Abelian theory \eqref{AB} for an arbitrary prime  $p$ using the explicit form of the genus $2$ averaged enumerator polynomial obtained in \cite{Dymarsky:2025agh}.
The result can be expressed as a statement about states in the genus $2$  Hilbert space $\H_2$,
\bea\label{WH}
{|{\rm W}_2\rangle\over N(n)}-{p^{n-1}+1\over p^{n-1}+p}{|{\rm W}_1\rangle \otimes |{\rm W}_1\rangle\over N(n)^2}= {p^{n-3}\over (p^{n-1}+1)(p^{n-2}+1)}\sum_{\gamma\in SL(2,\Z_p)} U_\gamma\, {\mathbb{P}} \in \H_2.
\eea
Here $|{\rm W}_{\sf g}\rangle$ are the same as in section \ref{sec:warmup}, which can be defined as
\bea
|{\rm W}_{\sf g}\rangle=\sum_{\C} |\C\rangle_{\sf g}\in \H_{\sf g},
\eea
and the states $|\C\rangle_{\sf g}$ are genus $\sf g$ full enumerators of codes $\C$, and in particular $\langle \Omega|\C\rangle_{\sf g}=W_{\sf g,\C}(\psi_{\alpha\beta})$ as defined in  
(\ref{averageC},\ref{averageC2}).
The sum in the RHS of \eqref{WH} is over the image of the modular group; its size is $p(p^2-1)$. Finally, ${\mathbb{P}}$ is the complex conjugation map from $\H_1^*$ to $\H_1$.

The same result can be rewritten in a more conventional form, making an explicit connection with the wormhole ansatz of \cite{Cotler:2020hgz},
\bea
\label{Whtau}
{\Wh}(\tau_1,\tau_2)={p^{2n-3}\over (p^{n-1}+1)(p^{n-2}+1)}\sum_{\gamma\in \Gamma(p)\backslash SL(2,\Z)} f_p(\tau_1,\gamma\, \tau_2) \\
-{p-1\over p^{n-1}+p}\avg{Z_{\sf CFT}(\tau_1)}\avg{Z_{\sf CFT}(\tau_2)}, \nonumber
\eea
where
\bea
\label{seed}
f_p=\left({1\over p}\sum_{\alpha\beta \in \Z_p} \psi_{\alpha\beta}(\tau_1) \psi^*_{\alpha\beta}(-\bar \tau_2)\right)^n. 
\eea
First term in \eqref{Whtau} is easy to interpret. Up to an overall coefficient  this is just the resolution of identity -- the naive TQFT amplitude on a cylinder $\Sigma_1 \times [0,1]$, which is subsequently summed over the modular group acting on one of the sides. Its form is in agreement with the interpretation of the  modular group  average $\sum_\gamma U_\gamma$ as the full non-perturbative ``Wheeler-DeWitt'' evolution operator of TQFT gravity, that projects on the Hilbert space of the alpha states \cite{Barbar:2025vvf}. Note, the result \eqref{Whtau} is exact, it includes the sum over all possible double-sided topologies ending on the product of two tori.

Second term goes beyond the ansatz of \cite{Cotler:2020hgz} and is more difficult to interpret. Yet it is necessary for consistency, e.g.~to ensure that $\Wh(\tau_1,\tau_2)$ vanish in the limit when the imaginary parts of both $\tau_1,\tau_2$ are taken to infinity.   

When $n\gg 1$ both terms in \eqref{Whtau} are exponentially suppressed as $e^{-O(n)}$. To obtain $\Wh$ in the case of \VTQFT\, gravity one should substitute $\psi_{\alpha\beta}$ by the Ising characters \eqref{substitution}. Here we assume that exponentially small corrections to \eqref{TPFlargec} and its genus two counterpart \eqref{seedPsdi0} combined will produce a subleading term relative to \eqref{Whtau}

As an interesting exercise we can consider fixed $n>2$ and take $p\rightarrow \infty$. In this limit we expect to recover wormhole the amplitude for the $U(1)$-gravity. For convenience we use normalization such that $\avg{Z_{\sf CFT}(\Omega_{\sf g}}$ for the whole Narain ensemble is given by the genus $\sf g$ real analytic Eisenstein series $E_{n/2}(\Omega_{\sf g})$, in other words $Z_{\sf CFT}(\tau)$ is understood as the partition function of $U(1)^n \times U(1)^n$ primaries. In the large $p$ limit we find 
\bea
\label{WhTau}
{\Wh}(\tau_1,\tau_2)=\lim_{p\rightarrow \infty}\sum_{\gamma\in \Gamma(p)\backslash SL(2,\Z)} f_p(\tau_1,\gamma\, \tau_2),
\eea
where (this expression is written assuming the so-called rigid embedding \cite{Aharony:2023zit}) 
\bea
\psi_{\alpha\beta}(\tau)=\Im(\tau)^{1/2}\sum_{n,m\in \Z} e^{i\pi \tau (pn+a+pm+b)^2/(2p)-i\pi \bar \tau (pn+a-pm-b)^2/(2p)}.
\eea
Taking the $p\rightarrow \infty$ limit readily gives the resolution of identity in $U(1)$-gravity,
\bea
\label{lim}
f_{U(1)}(\tau_1,\tau_2)=\lim_{p\rightarrow \infty} f_p(\tau_1,\tau_2)=\left({\Im(\tau_1)\Im(\tau_2)\over |\tau_1+\tau_2|^2}\right)^{n/2}. 
\eea
This leads to the following naive expression for the full wormhole amplitude \cite{Cotler:2020hgz},\footnote{This expression is twice as large as the proposal of \cite{Cotler:2020hgz} that only includes the sum over $PSL(2,\Z)$. In the context of gravity, this factor of $2$ was pointed out in \cite{Yan:2023rjh}.}
\bea
\label{naive}
\sum_{\gamma\in SL(2,\Z)} f_{U(1)}(\tau_1,\gamma\tau_2),
\eea
but it is not correct: a direct computation of \eqref{wh} in $U(1)$-gravity shows a mismatch \cite{Collier:2021rsn}.\footnote{The mismatch persists after taking the factor of $2$ into account. We thank the authors of \cite{Collier:2021rsn} for correspondence on this point.} 
We attribute that  to  the order of limits. While \eqref{lim} is correct for fixed $\tau_1,\tau_2$, the sum in \eqref{WhTau} includes terms with the imaginary part of $\gamma\,\tau_2$ being small,  when $f_p(\tau_1,\gamma\tau_2) \neq f(\tau_1,\gamma\,\tau_2)$ even for $p\rightarrow \infty$. It is interesting to note that numerically the difference between the exact and naive result \eqref{naive} is small. 

The lesson we can learn here is that the naive wormhole amplitude  in Virasoro TQFT with the seed given by the resolution of identity \cite{Cotler:2020hgz,Collier:2023fwi}
\bea
\label{amplitude}
f=Z_{\sf boson}(\tau_1) Z_{\sf boson}(\tau_2) \left({\Im(\tau_1)\Im(\tau_2)\over |\tau_1+\tau_2|^2}\right)^{n/2},\qquad n=1,
\eea
may acquire corrections due to necessary regularization, which is discussed in subsection \ref{sec:VTQFTeffective} below. 

We end with a speculation. In light of the simple form of the wormhole amplitude in \VTQFT\, gravity  given by the resolution of identity in the Abelian theory \eqref{seed}, we expect the seed of the gravity wormhole amplitude, given by \eqref{amplitude} with $n=2$ \cite{Cotler:2020ugk}, to be the resolution of identity in some effective theory.

\subsection{Holographic codes}
\label{sec:holcodes} 
There are many different codes  that already made their appearance in our calculations. Their main role was technical, to provide ways to enumerate different boundary CFTs, as well as TQFT states on various topologies. 

\begin{figure}
    \centering
    \includegraphics[width=0.5\linewidth]{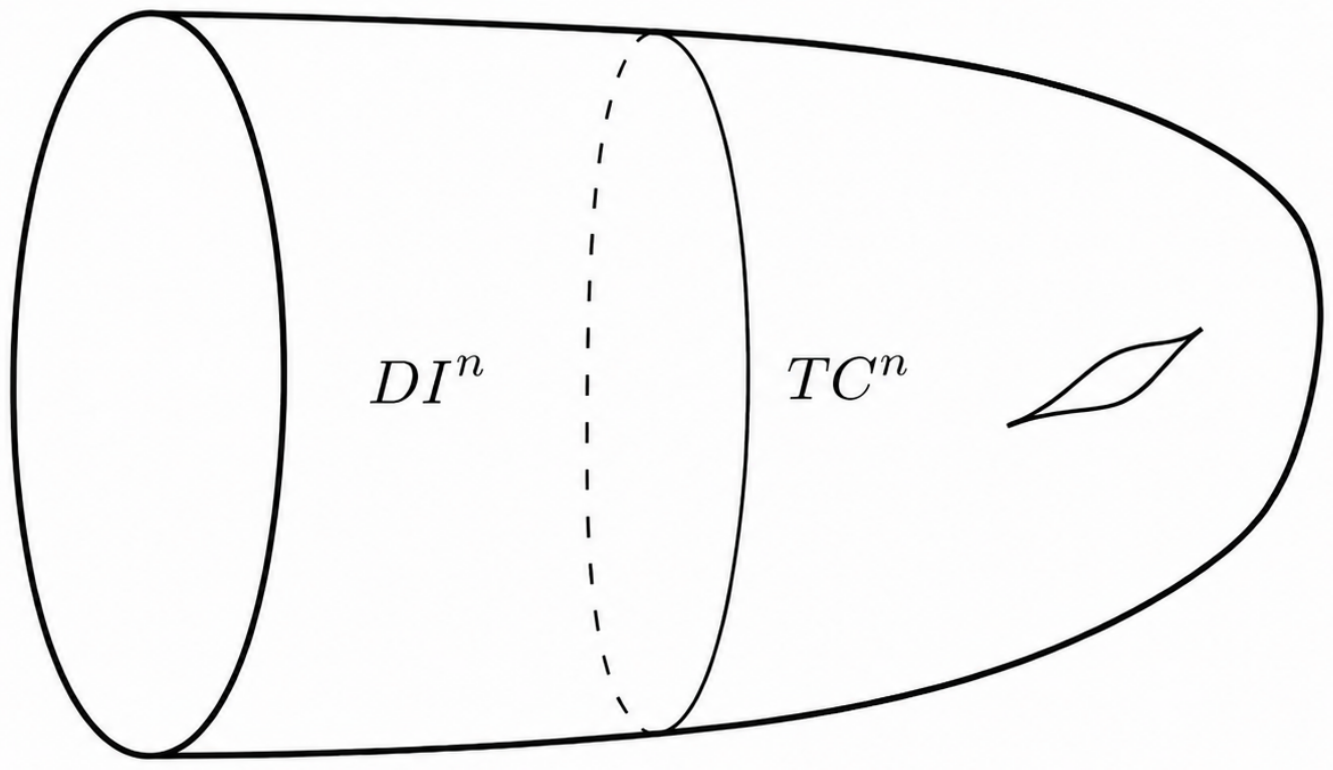}
    \caption{In the $c\gg 1$ limit the \VTQFT\, gravity condenses to an Abelian phase described by $n$ copies of the Toric Code summed over topologies. It is separated by an interface with the $n$ copies of the Doubled Ising, living near the boundary. The interface embeds the Hilbert space of the TC${}^n$  inside the Hilbert space of the DI${}^n$, providing a toy model of the holographic code.}
    \label{fig}
\end{figure}

The \VTQFT\, gravity  exhibits yet another type of code, a toy version of the holographic code of \cite{Almheiri:2014lwa}. As we discussed in section \ref{sec:largen},  in  the large central charge limit the \VTQFT\, gravity condenses to an Abelian phase, separated from the physical boundary by an interface, as shown in Fig.~\ref{fig}. The interface is a linear superposition of all $n!$ interfaces embedding the Hilbert space $\H_{\sf g}$ of the Abelian TC${}^n$ theory into the full Hilbert space $\H_{\sf g}$ of the DI${}^n$ theory. Only the Abelian TC${}^n$ theory should be summed over bulk topologies. The interface is assumed to be close to the boundary such that the DI${}^n$ always lives inside a ``slab'' $\Sigma_{\sf g}\times [0,1]$.
Since the theories in the bulk are topological, the interface can be moved to the boundary, providing a way to couple the Abelian theory directly to 2d CFTs. But the picture when the interface is close but finite distance away from the boundary is more physical and suggests the following interpretation. Near the boundary, in the UV region, bulk theory is more complicated and includes all states of the physical boundary. The interface connects it with some effective IR theory with a smaller Hilbert space. This is the ``low energy'' code subspace. In a more complicated model, one can imagine a cascade of condensations, providing an explicit connection with the RG picture outlined in \cite{Bao:2024ixc}.

\subsection{Virasoro TQFT is an effective theory}
\label{sec:VTQFTeffective}
Given that the prescription to sum over topologies \eqref{sumovertopologies} is universal, i.e.~TQFT independent, it is tempting to apply it to Virasoro TQFT to evaluate the partition function of the 3d quantum gravity. Here we are not specific about a particular formulation of the corresponding topological theory \cite{Kashaev:1998fc,Teschner:2010je,Collier:2023fwi,Hartman:2025ula}, as we expect any straightforward application of \eqref{sumovertopologies}  not to succeed. Indeed, as was pointed out in \cite{Collier:2023fwi} the partition function of the Virasoro TQFT on many off-shell topologies is ill-defined. One reason for that is easy to identify: Virasoro TQFT, that is closely related to  Chern-Simons theory with non-compact gauge group \cite{Witten:1988hc},  has a continuous spectrum leading to divergences.
Here a comparison with the Abelian case offers a helpful perspective. 

Let us consider the case of $U(1)$-gravity that is holographically dual to the ensemble of all Narain theories \cite{Afkhami-Jeddi:2020ezh,Maloney:2020nni}. The bulk theory is the Abelian Chern-Simons theory with the non-compact gauge group $\R^n \times \R^n$, placed on handlebody topologies, and 
completed with the prescription that only contributions of topologically trivial gauge fluctuations are taken into account. 
Beyond this essentially semiclassical definition the bulk theory is not well-defined \cite{Maloney:2020nni}. There is a related issue, the sum over handlebodies of a sufficiently large genus $g>n+1$ is divergent. Hence this theory is not amenable to the sum over topologies \eqref{sumovertopologies} that assumes  the intermediate $g\rightarrow \infty$ limit. The $U(1)$-gravity can be regularized by considering TQFT gravity based on an Abelian CS theory with a compact gauge group, e.g.~the AB theory \eqref{AB} coupled to Narain CFTs at the physical boundary. The non-compact theory then can be recovered in the $p\rightarrow \infty$ limit. Indeed, in this limit (for $n$>2) the discrete ensemble of Narain theories that is holographically dual to Abelian TQFT gravity will densely populate the whole Narain moduli space \cite{Aharony:2023zit,Barbar:2025krh}. As was discussed in section \ref{sec:warmup}, the Abelian theory with the compact gauge group can be summed over topologies leading to a well-defined result. Crucially, $p\rightarrow \infty$ and $g\rightarrow \infty$ limits do not commute. This is evident from the leading divergent part of the boundary ensemble-averaged genus $g>n+1$ partition function in the compact case \cite{Barbar:2025krh},
\bea
\avg{Z_{\rm CFT}(\Omega)} \sim p^{n(g+1-n)/2} E_{g}^{\mathcal O}(G+B), \qquad p\rightarrow \infty.
\eea
The leading divergence is independent of the period matrix $\Omega$. (The expression in the RHS  is the Eisenstein series of the {\it orthogonal} group; it  depends on the point on the Narain moduli space $G+B$ specified by the physical boundary state $\langle \Omega|$.) The upshot is that, while the regularized, well-defined bulk theory based on the compact  Abelian CS can be summed over topologies, the non-compact version is only defined  semiclassically on certain simple topologies. To rephrase it, $U(1)$-gravity is an effective theory which should be summed over certain but not all topologies, each corresponding to a whole infinite class of topologies in the compact case. 
This has direct parallels with the semiclassical quantum gravity and suggests the following picture.

Semiclassical quantum gravity, including its Virasoro TQFT formulation, is an effective theory; its sum over off-shell topologies is not well-defined. Instead semiclassical gravity can hypothetically be defined  as  a limit through a sequence of some well-defined ``regularized'' theories. Each of the regularized theories can be summed over all topologies as defined in \eqref{sumovertopologies}. Each saddle-point value of semiclassical gravity (Virasoro TQFT amplitude), evaluated on a particular on-shell topology, will correspond to a sum over an infinite class of topologies of the regularized theory. Such a  regularized theory  will be holographically dual to a well-defined boundary CFT or, more likely, an ensemble of  boundary  CFTs. 
We note that such a sequence of regularized theories, and hence the limit, if they exist, may not be unique.

One way to think about possible regularizations of quantum gravity is in terms of the modular bootstrap. Going back to Abelian case, the space of Narain theories can be defined as a space of solutions of the full set of modular bootstrap constraints for theories with $U(1)^n \times U(1)^n$ symmetry.
Then the finite ensemble related to compact  CS  can be defined by modifying the bootstrap constraints.  In particular S-invariance, which has the form of the  Fourier transform \cite{afkhami2019fast}, should be replaced by a discrete Fourier sum.  What is non-trivial here, the modified set of the bootstrap constraints admits many solutions for each value of the regulator $p$, that eventually can be taken to infinity to recover the whole space of Narain theories.
In the case of semiclassical gravity, the attempts to define the dual boundary ensemble by deforming or relaxing some of the modular bootstrap constraints \cite{Jafferis:2025vyp,Belin:2026pko} can be understood as introducing such regularization schemes.

\section{Summary and Outlook}
\label{sec:summary}
In this paper we considered a model of 3d quantum gravity defined within the framework of TQFT gravity  \cite{Barbar:2023ncl,Dymarsky:2024frx,Barbar:2025vvf}. Starting from $n+\bar n$ copies of $c=1/2$ Virasoro TQFT (3d Ising TQFT), we evaluated the sum over all bulk topologies with a fixed boundary. The bulk sum, defined in \eqref{sumovertopologies}, can be understood as an average over all generalized Heegaard splittings.  
The bulk theory thus defined is holographically dual to a weighted ensemble of all 2d CFTs whose chiral algebra is ${{\vphantom{\overline{\rm Vir}}{\rm Vir}}_{1/2}^{\otimes n} \times \overline{\rm Vir}_{1/2}^{\otimes \bar n}}$ or an extension thereof. Modulo technicalities genus $\sf g$ partition function of the bulk theory is given by the genus $\sf g$ averaged enumerator polynomial of the  $\D$-codes. The latter are some auxiliary linear binary codes introduced in the context of framed VOAs \cite{griess2001virasoro}. 

We provided a partial analytic argument establishing the holographic duality, i.e.~the equality between the bulk and boundary ensemble-averaged partition functions, 
for $n=\bar n<8$. It was supplemented by an explicit computer algebra check for $n=\bar n<5$. A proof for general $n,\bar n$ is left for future work \cite{inprogress}. 

Our model, which we refer to as \VTQFT\, gravity, further simplifies in the large central charge limit $n=\bar n=2c \gg 1$. In this limit its torus and higher genus partition functions admit explicit closed-form expressions.

The model of \VTQFT\, gravity offers an explicit and controllable setting to study the relation between the properties of the boundary ensemble and the sum over topologies. The bulk sum includes all possible topologies, but the bulk TQFT, $c=1/2$ Virasoro TQFT in our case, organizes then into equivalence classes. It would be interesting to understand if there is a canonical simple representative in each class such that the bulk sum could be rewritten to include only these representatives. Combined with full control over the boundary ensemble, it can be used to further develop the ideas of \cite{Belin:2026pko} that connect the OPE statistics to a restricted sum over a particular class of topologies.

When the central charge is large, the boundary ensemble is self-averaging.
This is a general expectation, confirmed in the models considered in this
work for $n\gg {\sf g}$. One can ask whether the sum over topologies can
be understood as a mean over a self-averaging ensemble, which can be
approximated with exponential precision by a typical representative~\cite{Cummings}.
For the Abelian model of section~\ref{sec:warmup} this is indeed the case,
since the sum over topologies can be rewritten as a sum over symplectic
codes $\mathcal{S}$~\cite{Dymarsky:2024frx}, which is self-averaging for
${\sf g}\gg n$. It would be interesting to understand whether this is a
general property of the bulk sum, and what makes a topology typical.

As a final remark, we note that the average over all generalized Heegaard splittings \eqref{sumovertopologies} deserves to be studied on its own, without reference to any particular TQFT $\T$. Although the size of the mapping class group is infinite, and therefore the expression in the  RHS of \eqref{sumovertopologies} is formal, one can define 
it probabilistically in terms of a random walk on the mapping class group \cite{Maher_2011}. 

\acknowledgments
I thank Ahmed Barbar, Aidan Herderschee, Juan Maldacena, Brandon Rayhaun, Alfred Shapere, and Herman Verlinde for discussions. This work is  supported by the IBM Einstein Fellow Fund and the NSF under grant 2310426.

\bibliographystyle{JHEP}
\bibliography{VTQFT}

\end{document}